\documentclass[12pt]{article}

\global\arraycolsep=1pt
\usepackage{graphicx}
\usepackage{amsmath}
\usepackage{amssymb}
\usepackage {pstricks}
\let\psgrid\relax
\newcommand{\FigureSize}{0.6}
\newcommand{\inverseFigureSize}{1.66667}

\newcommand{\be}{\begin{equation}}
\newcommand{\ee}{\end{equation}}
\newcommand{\beq}{\begin{equation}}
\newcommand{\eeq}{\end{equation}}
\newcommand{\ba}{\begin{eqnarray}}
\newcommand{\ea}{\end{eqnarray}}
\newcommand{\beqa}{\begin{eqnarray}}
\newcommand{\eeqa}{\end{eqnarray}}
\newcommand{\CR}{\nonumber \\}
\newcommand{\ds}{\displaystyle}

\newcommand{\fgref}[1]{Fig.\ \ref{#1}}
\newcommand{\bebox}[1]{
 \begin{equation}
  \fbox{
\rule[-.35cm]{0cm}{1cm}{~~~${\displaystyle{#1}}$~~~}}
}
\newcommand{\eebox}{\end{equation}}
\newcommand{\babox}[1]{
 \begin{equation}
  \fbox{\rule[-.5cm]{0cm}{1cm}{~~~${
   \begin{array}{rcl}  
    #1
   \end{array}
  }$~~~}}
}
\newcommand{\eabox}{\end{equation}}
\renewcommand{\theequation}{\thesection.\arabic{equation}}
\newcommand{\Section}{\setcounter{equation}{0} \section}
\renewcommand{\thefootnote}{\fnsymbol{footnote}}
\renewcommand{\a}{\alpha}
\renewcommand{\b}{\beta}
\newcommand{\g}{\gamma}
\newcommand{\ve}{\varepsilon} 
\newcommand{\la}{\lambda}
\newcommand{\lam}{\lambda}

\newcommand{\cO}{{\cal O}}
\newcommand{\bC}{{\mathbb C}}
\newcommand{\bN}{{\mathbb N}}
\newcommand{\bZ}{{\mathbb Z}}
\newcommand{\mba}{{\mathbf a}}     
\newcommand{\mbq}{{\mathbf q}}            
\newcommand{\mbt}{{\mathbf t}}                   
\newcommand{\proof}{\noindent{\it Proof.\hskip10pt}} 
\newcommand{\qed}{\hfill\fbox{}}
\renewcommand{\(}{\left(}
\renewcommand{\)}{\right)}
\newcommand{\ha}{{1\over2}}
\newcommand{\Exp}[1]{\exp\left\{#1\right\}}
\newcommand{\bu}{\bullet}    

\newcommand{\Smat}[1]{\left[\begin{matrix}#1\end{matrix}\right]}
\newcommand{\Diff}[1]{T_{q,x_{#1}}}

\newcommand{\tz}{\tilde z}

\newcommand{\tQ}{\widetilde Q}

\newcommand{\Ciio}[5]{C_{#1#2}{}^{#3}(#4,#5)}
\newcommand{\Cooi}[5]{C^{#1#2}{}_{#3}(#4,#5)}
\newcommand{\Cioo}[5]{C_{#1}{}^{#2#3}(#4,#5)}
\newcommand{\Coii}[5]{C^{#1}{}_{#2#3}(#4,#5)}
\newcommand{\Ya}[2]{{ #1_{#2} }}
\newcommand{\Yav}[2]{{ #1_{#2}{}^\vee }}
\newcommand{\Zm}[2]{Z^{\mathrm {#1}}_{#2}}
\newcommand{\fla}[3]{f_{#1}\left(#2,#3\right)}
\newcommand{\unitbox}
{\setlength{\unitlength}{0.5pt}
\begin{picture}(10,10)
\put(0,10){\line(1,0){10}}
\put(0,0){\line(1,0){10}}
\put(0,0){\line(0,1){10}}
\put(10,0){\line(0,1){10}}
\end{picture}}
\newcommand\HopfS{%
\psline(-1,3)(-1,1)
\psline(-3,1)(-1,1)
\psline      (-1,1)(1,-1)
\psline            (1,-1)(3,-1)
\psline            (1,-1)(1,-3)
\rput(-3.5,1){$\lambda$}
\rput(1,-3.5){$\mu$}
}
\newcommand\HopfSmiddle{%
\rput(0,0){\HopfS}
\psline[linewidth=2 pt](-.2,-.2)(.2,.2)
}

\newcommand\HopfSlr{%
\rput(0,0){\HopfS}
\psline[linewidth=2 pt]( 2,-.8)( 2,-1.2)
\psline[linewidth=2 pt](-2, .8)(-2, 1.2)
}

\newcommand\HopfT{%
\psline              (1,1)(1,3)
\psline              (1,1)(3,1)
\psline       (-1,-1)(1,1)
\psline(-3,-1)(-1,-1)
\psline(-1,-3)(-1,-1)
\rput(-3.5,-1){$\lambda$}
\rput(-1,-3.5){$\mu$}
}

\newcommand\HopfTlr{%
\rput(0,0){\HopfT}
\psline[linewidth=2 pt](-2,-.8)(-2,-1.2)
\psline[linewidth=2 pt]( 2, .8)( 2, 1.2)
}

\newcommand\FigSlicingFlop{%
\begin{figure}[h]
\psset{unit=\FigureSize cm}
\begin{center}
\begin{pspicture}(-13,-3)(13,3)
\psgrid
\rput(-10, 0)\HopfSmiddle
\rput(  0, 0)\HopfSlr
\rput( 10, 0)\HopfTlr
\psline[arrowsize=5pt]{->}(-6,0)(-4,0)
\rput(-5,-1){slicing}
\psline[arrowsize=5pt]{->}(4,0)(6, 0)
\rput(5,-1){flop}
\end{pspicture}
\end{center}
\caption{Slicing and Flop; the preferred direction is indicated by the bold line
\hbox{\vrule height6pt width2pt}
.}
\end{figure}
}

\newdimen\Sx 
\newdimen\Sy
\newdimen\Ex 
\newdimen\Ey
\newdimen\Mx 
\newdimen\My
\newdimen\Tx 
\newdimen\Ty

\newdimen\Fx 
\newdimen\Fy
\newdimen\Bx 
\newdimen\By
\newdimen\Rx 
\newdimen\Ry
\newdimen\Lx 
\newdimen\Ly
\newdimen\Ux 
\newdimen\Uy
\newdimen\Dx 
\newdimen\Dy

\newdimen\txtlen
\newdimen\txthgt

\newcommand{\WhiteTriangle}[4]{
\psline(0,0)(#1,#2)
\psline(0,0)(#3,#4)
\psline[linewidth=\Lwd pt](#1,#2)(#3,#4)
}
\newcommand{\AAx}{0.4131714875}
\newcommand{\AAy}{0.1711412338}
\newcommand{\BBx}{0.4131714875}
\newcommand{\BBy}{0.1711412338}
\newcommand{\Lwd}{0.5}
\newcommand{\triangleW}{
\WhiteTriangle{\AAx}{-\AAy}{\AAx}{\AAy}
}
\newcommand{\triangleSW}{
\WhiteTriangle{ \BBx}{\BBy}{\BBy}{\BBx}
}
\newcommand{\triangleS}{
\WhiteTriangle{\AAy}{\AAx}{-\AAy}{\AAx}
}
\newcommand{\triangleSE}{
\WhiteTriangle{-\BBy}{\BBx}{-\BBx}{\BBy}
}
\newcommand{\triangleE}{
\WhiteTriangle{-\AAx}{\AAy}{-\AAx}{-\AAy}
}
\newcommand{\triangleNE}{
\WhiteTriangle{-\BBx}{-\BBy}{-\BBy}{-\BBx}
}
\newcommand{\triangleN}{
\WhiteTriangle{-\AAy}{-\AAx}{\AAy}{-\AAx}
}
\newcommand{\triangleNW}{
\WhiteTriangle{\BBy}{-\BBx}{\BBx}{-\BBy}
}

\makeatletter
\def\CArrow{\@ifnextchar[{\@CArrow}{\@CArrow[]}} 
\def\@CArrow[#1]#2#3#4#5#6{{
\advance \Sx by #2\p@
\advance \Sy by #3\p@
\advance \Ex by #4\p@
\advance \Ey by #5\p@
\advance \Mx by 0.5\Sx
\advance \Mx by 0.5\Ex
\advance \My by 0.5\Sy
\advance \My by 0.5\Ey
\advance \Tx by 0.4\Sx
\advance \Tx by 0.6\Ex
\advance \Ty by 0.4\Sy
\advance \Ty by 0.6\Ey
\setbox0\hbox{#6}\advance\txtlen by 0.035146\wd0%
\setbox0\vbox{#6}\advance\txthgt by 0.035146\ht0%
\txtlen=\inverseFigureSize\txtlen
\txthgt=\inverseFigureSize\txthgt
\psline
(\expandafter\Rval\the\Mx,\expandafter\Rval\the\My)
(\expandafter\Rval\the\Ex,\expandafter\Rval\the\Ey)%
\@tfor\opti :=#1\do{
\if\opti O 
                \psline[arrowsize=5pt]
                {->}(\expandafter\Rval\the\Sx,\expandafter\Rval\the\Sy)
                (\expandafter\Rval\the\Tx,\expandafter\Rval\the\Ty)%
\fi
\if\opti W 
                \psline
                (\expandafter\Rval\the\Sx,\expandafter\Rval\the\Sy)
                (\expandafter\Rval\the\Tx,\expandafter\Rval\the\Ty)%
              \rput(\expandafter\Rval\the\Tx,\expandafter\Rval\the\Ty){\triangleW}%
\fi
\if\opti w 
                \psline
                (\expandafter\Rval\the\Sx,\expandafter\Rval\the\Sy)
                (\expandafter\Rval\the\Tx,\expandafter\Rval\the\Ty)%
              \rput(\expandafter\Rval\the\Tx,\expandafter\Rval\the\Ty){\triangleSW}%
\fi
\if\opti S 
                \psline
                (\expandafter\Rval\the\Sx,\expandafter\Rval\the\Sy)
                (\expandafter\Rval\the\Tx,\expandafter\Rval\the\Ty)%
              \rput(\expandafter\Rval\the\Tx,\expandafter\Rval\the\Ty){\triangleS}%
\fi
\if\opti s 
                \psline
                (\expandafter\Rval\the\Sx,\expandafter\Rval\the\Sy)
                (\expandafter\Rval\the\Tx,\expandafter\Rval\the\Ty)%
              \rput(\expandafter\Rval\the\Tx,\expandafter\Rval\the\Ty){\triangleSE}%
\fi
\if\opti E 
                \psline
                (\expandafter\Rval\the\Sx,\expandafter\Rval\the\Sy)
                (\expandafter\Rval\the\Tx,\expandafter\Rval\the\Ty)%
              \rput(\expandafter\Rval\the\Tx,\expandafter\Rval\the\Ty){\triangleE}%
\fi
\if\opti e 
                \psline
                (\expandafter\Rval\the\Sx,\expandafter\Rval\the\Sy)
                (\expandafter\Rval\the\Tx,\expandafter\Rval\the\Ty)%
              \rput(\expandafter\Rval\the\Tx,\expandafter\Rval\the\Ty){\triangleNE}%
\fi
\if\opti N 
                \psline
                (\expandafter\Rval\the\Sx,\expandafter\Rval\the\Sy)
                (\expandafter\Rval\the\Tx,\expandafter\Rval\the\Ty)%
              \rput(\expandafter\Rval\the\Tx,\expandafter\Rval\the\Ty){\triangleN}%
\fi
\if\opti n 
                \psline
                (\expandafter\Rval\the\Sx,\expandafter\Rval\the\Sy)
                (\expandafter\Rval\the\Tx,\expandafter\Rval\the\Ty)%
              \rput(\expandafter\Rval\the\Tx,\expandafter\Rval\the\Ty){\triangleNW}%
\fi
\if\opti F 
        \advance \Fx by  -0.2\Sx
        \advance \Fx by   1.2\Ex
        \advance \Fy by -0.2\Sy
        \advance \Fy by  1.2\Ey
        \rput(\expandafter\Rval\the\Fx,\expandafter\Rval\the\Fy){#6}
        \advance \Fx by -\Fx
        \advance \Fy by -\Fy
\fi
\if\opti B 
        \advance \Bx by   1.2\Sx
        \advance \Bx by  -0.2\Ex
        \advance \By by  1.2\Sy
        \advance \By by -0.2\Ey
        \rput(\expandafter\Rval\the\Bx,\expandafter\Rval\the\By){#6}
        \advance \Bx by -\Bx
        \advance \By by -\By
\fi
\if\opti R 
        \advance \Rx by \Mx
        \advance \Rx by 0.5\txtlen
        \advance \Rx by \txthgt
        \advance \Ry by \My
        \rput(\expandafter\Rval\the\Rx,\expandafter\Rval\the\Ry){#6}
        \advance \Rx by -\Rx
        \advance \Ry by -\Ry
\fi
\if\opti L 
        \advance \Lx by \Mx
        \advance \Lx by -0.5\txtlen
        \advance \Lx by -\txthgt
        \advance \Ly by \My
        \rput(\expandafter\Rval\the\Lx,\expandafter\Rval\the\Ly){#6}
        \advance \Lx by -\Lx
        \advance \Ly by -\Ly
\fi
\if\opti U 
        \advance \Ux by \Mx
        \advance \Uy by \My
        \advance \Uy by 0.5\txthgt
        \advance \Uy by \txthgt
        \rput(\expandafter\Rval\the\Ux,\expandafter\Rval\the\Uy){#6}
        \advance \Ux by -\Ux
        \advance \Uy by -\Uy
\fi
\if\opti D 
        \advance \Dx by \Mx
        \advance \Dy by \My
        \advance \Dy by -0.5\txthgt
        \advance \Dy by -\txthgt
        \rput(\expandafter\Rval\the\Dx,\expandafter\Rval\the\Dy){#6}
        \advance \Dx by -\Dx
        \advance \Dy by -\Dy
\fi
}
\advance \Sx by -\Sx
\advance \Sy by -\Sy
\advance \Ex by -\Ex
\advance \Ey by -\Ey
\advance \Mx by -\Mx
\advance \My by -\My
\advance \Tx by -\Tx
\advance \Ty by -\Ty
\advance\txtlen by -\txtlen
\advance\txthgt by -\txthgt
}}
{\catcode`\p=12\catcode`\t=12\gdef\Rval#1pt{#1}}
\makeatother

\makeatletter
\def\LPut{\@ifnextchar[{\@LPut}{\@LPut[]}} 
\def\@LPut[#1]#2#3#4#5#6{{
\advance \Sx by #2\p@
\advance \Sy by #3\p@
\advance \Ex by #4\p@
\advance \Ey by #5\p@
\advance \Mx by 0.5\Sx
\advance \Mx by 0.5\Ex
\advance \My by 0.5\Sy
\advance \My by 0.5\Ey
\setbox0\hbox{#6}\advance\txtlen by 0.035146\wd0%
\setbox0\vbox{#6}\advance\txthgt by 0.035146\ht0%
\txtlen=\inverseFigureSize\txtlen
\txthgt=\inverseFigureSize\txthgt
\@tfor\opti :=#1\do{
\if\opti F 
        \advance \Fx by  -0.2\Sx
        \advance \Fx by   1.2\Ex
        \advance \Fy by -0.2\Sy
        \advance \Fy by  1.2\Ey
        \rput(\expandafter\Rval\the\Fx,\expandafter\Rval\the\Fy){#6}
        \advance \Fx by -\Fx
        \advance \Fy by -\Fy
\fi
\if\opti B 
        \advance \Bx by   1.2\Sx
        \advance \Bx by  -0.2\Ex
        \advance \By by  1.2\Sy
        \advance \By by -0.2\Ey
        \rput(\expandafter\Rval\the\Bx,\expandafter\Rval\the\By){#6}
        \advance \Bx by -\Bx
        \advance \By by -\By
\fi
\if\opti R 
        \advance \Rx by \Mx
        \advance \Rx by 0.5\txtlen
        \advance \Rx by \txthgt
        \advance \Ry by \My
        \rput(\expandafter\Rval\the\Rx,\expandafter\Rval\the\Ry){#6}
        \advance \Rx by -\Rx
        \advance \Ry by -\Ry
\fi
\if\opti L 
        \advance \Lx by \Mx
        \advance \Lx by -0.5\txtlen
        \advance \Lx by -\txthgt
        \advance \Ly by \My
        \rput(\expandafter\Rval\the\Lx,\expandafter\Rval\the\Ly){#6}
        \advance \Lx by -\Lx
        \advance \Ly by -\Ly
\fi
\if\opti U 
        \advance \Ux by \Mx
        \advance \Uy by \My
        \advance \Uy by 0.5\txthgt
        \advance \Uy by \txthgt
        \rput(\expandafter\Rval\the\Ux,\expandafter\Rval\the\Uy){#6}
        \advance \Ux by -\Ux
        \advance \Uy by -\Uy
\fi
\if\opti D 
        \advance \Dx by \Mx
        \advance \Dy by \My
        \advance \Dy by -0.5\txthgt
        \advance \Dy by -\txthgt
        \rput(\expandafter\Rval\the\Dx,\expandafter\Rval\the\Dy){#6}
        \advance \Dx by -\Dx
        \advance \Dy by -\Dy
\fi
}
\advance \Sx by -\Sx
\advance \Sy by -\Sy
\advance \Ex by -\Ex
\advance \Ey by -\Ey
\advance \Mx by -\Mx
\advance \My by -\My
\advance\txtlen by -\txtlen
\advance\txthgt by -\txthgt
}}
{\catcode`\p=12\catcode`\t=12\gdef\Rval#1pt{#1}}
\makeatother

\newcommand\FigUN{%
\begin{figure}[h]
\psset{unit=\FigureSize cm}
\begin{center}
\begin{pspicture}(0,0)(22,18)
\psgrid
\rput(0,8){%
\CArrow[OB]08 28{$\bu$}\CArrow[SB]2{10}28{$\Ya\la {2N}$}%
                       \CArrow[OL]28 46{$\Ya\la 1$}%
\CArrow[OB]24 44{$\bu$}\CArrow[SL]46 44{$\Ya\la 2$}\CArrow[OF]46 66{$\bu$}%
                       \CArrow[OL]44 62{$\Ya\la 3$}%
                       \CArrow[SL]62 60{$\Ya\la 4$}\CArrow[OF]62 82{$\bu$}%
\LPut[R]28 46{$Q_1$}%
\LPut[R]46 44{$Q_2$}%
\LPut[R]44 62{$Q_3$}%
\LPut[R]62 60{$Q_4$}%
}%
\rput(6,0){%
\psline[linestyle=dashed,dash=3pt 2pt](0,8)(2,6)%
\CArrow[OB]04 24{$\bu$}\CArrow[SL]26 24{$\Ya\la {2N-2}$}%
                       \CArrow[OL]24 42{$\Ya\la {2N-1}$}%
                       \CArrow[SF]42 40{$\Ya\la {2N  }$}\CArrow[OF]42 62{$\bu$}%
\LPut[R]26 24{$Q_{2N-2}$}%
\LPut[R]24 42{$Q_{2N-1}$}%
\LPut[R]42 40{$Q_{2N  }$}%
\rput(2,-1){$Z_L$}%
}%
\rput(10,8){%
\CArrow[EB]08 28{$\bu$}\CArrow[OB]2{10}28{$\Ya\la {2N}$}%
                       \CArrow[OL]28 46{$\Ya\la 1$}%
\CArrow[EB]24 44{$\bu$}\CArrow[OL]46 44{$\Ya\la 2$}\CArrow[EF]46 66{$\bu$}%
                       \CArrow[OL]44 62{$\Ya\la 3$}%
                       \CArrow[OL]62 60{$\Ya\la 4$}\CArrow[EF]62 82{$\bu$}%
\LPut[R]28 46{$Q_1$}%
\LPut[R]46 44{$Q_2$}%
\LPut[R]44 62{$Q_3$}%
\LPut[R]62 60{$Q_4$}%
}%
\rput(16,0){%
\psline[linestyle=dashed,dash=3pt 2pt](0,8)(2,6)%
\CArrow[EB]04 24{$\bu$}\CArrow[OL]26 24{$\Ya\la {2N-2}$}%
                       \CArrow[OL]24 42{$\Ya\la {2N-1}$}%
                       \CArrow[OF]42 40{$\Ya\la {2N  }$}\CArrow[EF]42 62{$\bu$}%
\LPut[R]26 24{$Q_{2N-2}$}%
\LPut[R]24 42{$Q_{2N-1}$}%
\LPut[R]42 40{$Q_{2N  }$}%
\rput(2,-1){$Z_R$}%
}%
\end{pspicture}
\end{center}
\caption{Choice of the preferred direction for $U(1)^N$ theory
}
\label{fg:UN}
\end{figure}
}

\newcommand\FigUone{%
\begin{figure}[h]
\psset{unit=\FigureSize cm}
\begin{center}
\begin{pspicture}(0,0)(14,7)
\psgrid
\rput(0,0){%
                       \CArrow[OB]06 24{$\nu$}%
\CArrow[OB]02 22{$\bu$}\CArrow[SL]24 22{$\lambda$}\CArrow[OF]24 44{$\bu$}%
                       \CArrow[OF]22 40{$\nu$}%
\LPut[R]06 24{$Q$}%
\LPut[R]24 22{$\Lambda$}%
\rput(2,-1){$Z_L$}%
}%
\rput(10,0){%
                       \CArrow[OB]06 24{$\nu$}%
\CArrow[EB]02 22{$\bu$}\CArrow[OL]24 22{$\lambda$}\CArrow[EF]24 44{$\bu$}%
                       \CArrow[OF]22 40{$\nu$}%
\LPut[R]06 24{$Q$}%
\LPut[R]24 22{$\Lambda$}%
\rput(2,-1){$Z_R$}%
}%
\end{pspicture}
\end{center}
\caption{Change of the preferred direction, 
which is indicated by the white arrow.
}
\label{fg:Uone}
\end{figure}
}

\begin{document}

%
\begin{titlepage}
\vspace{0.5cm}
\begin{center}
{\Large \bf
Changing the preferred direction  \\
of the refined topological vertex}
\vskip1.0cm
{\large Hidetoshi Awata and Hiroaki Kanno}
\vskip 1.0em
{\it 
Graduate School of Mathematics \\
Nagoya University, Nagoya, 464-8602, Japan}
\end{center}
\vskip1.5cm

\begin{abstract}
We consider the issue of the slice invariance
of refined topological string amplitudes, which means 
that they are independent of the choice of the preferred direction 
of the refined topological vertex. We work out two examples. 
The first example is a geometric engineering of five-dimensional 
$U(1)$ gauge theory with a matter. The slice invariance
follows from a highly non-trivial combinatorial identity which equates
two known ways of computing
the $\chi_y$ genus of the Hilbert scheme of points on ${\mathbb C}^2$.
The second example is concerned with the proposal that 
the superpolynomials of the colored Hopf link are
obtained from a refinement of topological open string amplitudes. 
We provide a closed formula for the superpolynomial, which 
confirms  the slice invariance when the Hopf link is colored with
totally anti-symmetric representations. 
However, we observe a breakdown of the slice invariance for other representations. 
\end{abstract}
\end{titlepage}


\renewcommand{\thefootnote}{\arabic{footnote}} \setcounter{footnote}{0}



\Section{Introduction}



All genus topological string amplitudes on local toric Calabi-Yau 3-fold can be computed
by a diagrammatic rule, in terms of the topological vertex \cite{AMV, AKMV, ORV};
\beq
C_{\mu \lambda \nu} (q) = q^{\frac{\kappa(\nu)}{2}} s_{\lambda}(q^\rho) 
\sum_{\eta} s_{\mu/\eta}(q^{\lambda^\vee+\rho}) s_{\nu^\vee/\eta}(q^{\lambda+\rho})~, \label{tv}
\eeq
with three $U(\infty)$ representations, or partitions $\mu, \nu$ and $\lambda$, which will be identified 
as Young diagrams throughout the present paper. In \eqref{tv}
$s_{\lambda/\mu}(x)$ is the (skew) Schur function and 
$q^{\lambda+\rho}$ means the substitution $x_i := q^{\lambda_i -i + \frac{1}{2}}$, where $q$ is
related to the genus expansion parameter $g_s$ of topological string by $q=e^{- g_s}$. 
The dual partition $\lambda^\vee$ is defined by the transpose of the corresponding Young diagram. 
The index $\kappa(\nu)$ is related to the (relative) framing of the topological vertex.
The relation of Nekrasov's partition function \cite{Nek,NO,NY} to the topological string amplitudes
motivates us to seek a refinement of the topological vertex \cite{HIV}.
We first proposed such a refinement in \cite{AK1} by employing 
the (skew) Macdonald function $P_{\lambda/\mu}(x;q,t)$.
Later it has been improved by incorporating the framing factor \cite{Taki} as follows:
\begin{align}
{ \Ciio\mu\lambda\nu qt }
= 
 { \fla\nu qt }^{-1}
P_{\lambda}(t^{\rho}; q,t) 
\sum_{\eta} \left(\frac{q}{t}\right)^{\frac{|\eta|-|\nu|}{2}} 
~{\iota} P_{\mu^\vee/\eta^\vee}( -t^{\lambda^\vee}q^{\rho}; t,q)
P_{\nu/\eta}(q^{\lambda} t^{\rho}; q,t)~, 
\label{ourvertex}
\end{align}
where $q^{\lambda} t^{\rho}$ {\it etc.} means 
the specialization $x_i := q^{\lambda_i} t^{\frac{1}{2} - i}$ and
${ \fla\nu qt }$ is the framing factor. See \cite{AK2} for more details on notations.
When $q=t$, \eqref{ourvertex} reduces to the unrefined vertex \eqref{tv}.
A slightly different version of the refined topological vertex was introduced in \cite{IKV};
\begin{align}
C_{\mu\nu\lambda}^{(IKV)}(q_1, q_2) 
&=  
\left(\frac{q_1}{q_2}\right)^{\frac{||\nu||^2 + ||\lambda||^2}{2}} q_2^{\frac{\kappa(\nu)}{2}}
P_{\lambda^\vee}(q_2^{-\rho}; q_1, q_2)  \CR
&~~~\times \sum_{\eta} \left(\frac{q_1}{q_2} \right)^{\frac{|\eta|+|\mu|-|\nu|}{2}}
s_{\mu^\vee/\eta}(q_1^{-\lambda} q_2^{-\rho}) 
s_{\nu/\eta}(q_2^{-\lambda^\vee} q_1^{-\rho})~,
\end{align}
which allows an interesting interpretation in terms of  \lq\lq unisotropic\rq\rq\  plane partitions.
This is also related to the statistical model of melting crystal \cite{ORV, IKV}. 
It has been proposed that a certain topological open string amplitude 
computed from  $C_{\mu\nu\lambda}^{(IKV)}(q_1, q_2)$ gives homological 
invariants of the Hopf link \cite{GIKV}. The relation of the refined topological
vertex and homological invariants is one of the main subjects of this paper.

Though it is not manifest in the expression \eqref{tv},
the topological vertex is symmetric under the cyclic permutation of three partitions. 
However, it seems impossible to keep the cyclic symmetry for the refinements. 
Consequently, both ${ \Ciio\mu\lambda\nu qt }$
and $C_{\mu\nu\lambda}^{(IKV)}(q_1, q_2)$ have
a preferred direction, which gives us
an issue of the choice of the preferred direction in the rule of
diagrammatic computation by the refined topological vertex. 
In \cite{AK2} we have proved that our refined topological vertex gives
a building block of Nekrasov's partition function, where the parameters $q$ and $t$ 
are related the $\Omega$-background of Nekrasov by 
$(q,t) = (e^{\epsilon_2}, e^{-\epsilon_1})$.
To obtain Nekrasov's partition function we consider toric diagrams 
of the geometric engineering of ${\cal N}=2$ supersymmetric Yang-Mills
theory \cite{IK-P1, IK-P2, EK1, EK2, HIV}. For such toric diagrams
we required that the preferred direction at each vertex should be parallel each other.
This requirement largely restricts a possible choice 
of the preferred direction. However, for a certain diagram there are 
more than one choice. In this paper we consider the possibility of changing 
the preferred direction. It is often the case that a summation over
partitions remains for one choice of the preferred direction, while
we can compute the summation completely for the other choice.
Thus the independence of the preferred direction gives a kind of 
summation formula for partitions. In this paper we consider two
examples of such summation formulae (see \eqref{5Dconjecture} and 
\eqref{main} below). The slice invariance is important not only 
for the consistency of the formalism but also for practical computations. 

The first example is a topological closed string amplitude that appears 
in the geometric engineering of five-dimensional $U(1)$ theory with a matter. 
When we compute the closed string amplitude 
based on the refined topological vertex,
there are alternative choices of the preferred direction
in the toric diagram of the five-dimensional $U(1)$ theory. 
The agreement of the amplitude requires a highly non-trivial
combinatorial identity; 
\beqa
&&
\sum_{\lambda}\Lambda^{|\lambda|} 
\prod_{s\in\lambda}
\frac
{1-Q q^{ a(s)  } t^{ \ell(s)+1}}{{1-  q^{ a(s)  }} t^{ \ell(s)+1}}
\frac
{1-Q q^{-a(s)-1} t^{-\ell(s)  }} {{1-q^{-a(s)-1}} t^{-\ell(s) }}\CR
&&~~=
\exp\left\{\sum_{n>0} 
\frac{1}{n}
\frac{\Lambda^n}{1-\Lambda^n Q^n}
\frac{(1- t^n Q^n)(1- q^{-n} Q^n)}
{(1- t^{n})(1- q^{-n})}
\right\}, \label{5Dconjecture}
\eeqa
where the light hand side is a summation over partitions $\lambda$.  $a(s)$ and $\ell(s)$
in the product are the arm length and the leg length that are defined by
the corresponding Young diagram. 
The identity \eqref{5Dconjecture} was discussed before in \cite{IKS, PS, AK2}.
It is known that this amplitude gives the generating function of 
the equivariant $\chi_y$ genus of the Hilbert scheme of points on ${\mathbb C}^2$.
The validity of \eqref{5Dconjecture} has a deep geometrical meaning. Namely 
the left hand side of \eqref{5Dconjecture} comes 
from a computation of the generating function by the localization for a toric action 
on the Hilbert scheme which is induced by $(z_1, z_2) \to (e^{i\epsilon_1} z_1, e^{i\epsilon_2} z_2)$ 
on ${\mathbb C}^2$, where the fixed points are labeled by Young diagrams. 
On the other hand a formula that computes the $\chi_y$ genus of the Hilbert scheme 
of a surface $S$ from the $\chi_y$ genus of the underlying surface $S$ has been established in \cite{EGL}.
There is also a proposal of more general formula of the elliptic genus based on string theory
or the orbifold conformal field theory \cite{DMVV}. The right hand side of \eqref{5Dconjecture}
can be derived, if we apply these formulae to the equivariant $\chi_y$ genus of ${\mathbb C}^2$.
Thus the identity \eqref{5Dconjecture} equates these two ways of computing
the equivariant $\chi_y$ genus of  the Hilbert scheme, one by the localization principle, 
the other from the corresponding $\chi_y$ genus of the underlying manifold.

The second example comes from a refinement of topological open string amplitudes, 
which is concerned with the relation of the refined topological vertex
to homological refinements of the link polynomial. 
According to the proposal in \cite{GIKV}, the homological invariant of the Hopf link colored
with representations $\lambda$ and $\mu$ is proportional to 
\beq
Z_{\lambda,\mu}(q_1,q_2,Q) =
\sum_{\eta} (-Q)^{|\eta|} q_2^{\frac{1}{2}||\eta^\vee||^2} q_1^{\frac{1}{2}||\eta||^2}
\tilde Z_{\eta^\vee}(q_1,q_2) \tilde Z_{\eta}(q_2,q_1)
s_\lambda(q_1^{-\eta} q_2^{-\rho}) s_\mu(q_1^{-\eta} q_2^{-\rho})~, \label{GIKVsum}
\eeq
where
\beq
\tilde Z_\eta (q_2, q_1) = \prod_{(i,j) \in \eta} \left(1 - q_1^{\eta_j^\vee -i +1} q_2^{\eta_i -j} \right)^{-1}
= \prod_{s \in \eta} \left(1 - q_1^{\ell(s)+1} q_2^{a(s)} \right)^{-1}~.
\eeq
Note that the summation over the partition $\eta$ remains in the
definition of $Z_{\lambda,\mu}$. Therefore the fact that 
the homological invariants are polynomials in $Q$ is
not clear at all in the above proposal. In this paper we work out
the summation in \eqref{GIKVsum}, when both $\lambda$ and $\mu$ are
totally antisymmetric representations, $(\lambda, \mu) = (1^r, 1^s)$ and prove
\beq
\frac{Z_{1^r, 1^s}(Q; q,t)}{Z_{\bullet, \bullet}(Q; q,t)}
= (-1)^s t^{-\frac{1}{2}s(s-1)} e_r (t^\rho) 
e_s(\sqrt{\frac{q}{t}} Q q^{(1^r)} t^\rho, t^{-\rho})
\prod_{i=1}^r \left( 1- Q q^{\frac{1}{2}} t^{-i+\frac{1}{2}} \right)~, \label{main}
\eeq
where $(q,t)=(q_1^{-1}, q_2^{-1})$ and $\bullet$ means the trivial representation. 
$e_r(x)$ and $e_s(x,y)$ are the elementary symmetric functions.
The fundamental summation formula on the space of symmetric functions is the Cauchy formula \cite{Mac}.
However, we cannot apply the Cauchy formula directly to \eqref{GIKVsum}.
We will heavily rely on many properties of the Macdonald
functions and the Macdonald operator to be introduced
in section 4, so that we can use the Cauchy formula to
perform the summation over partitions. Though our proof of the formula 
\eqref{main} can be made without referring to the refined topological vertex, 
it is closely related to a change of the preferred direction
as has been argued by Taki \cite{Taki2}, (see section 3 for the explicit form of his conjecture).
The above formula agrees with his conjecture (up to normalization), 
when both $\lambda$ and $\mu$ are totally anti-symmetric representations. 
This means that we can show the invariance under the change of the preferred direction in this case.
However, for general representations Taki's conjecture should be appropriately
modified as discussed in section 6.
Recently the homological invariants for the pair $(1^k, 1)$ is constructed
by Yonezawa \cite{Yone} by the method of matrix factorization. If we apply his result
to the Hopf link, we find a complete agreement to our formula \eqref{main} up to an overall normalization.

The paper is organized as follows;
in section 2 we consider the toric diagram of the geometric engineering of 
five-dimensional $U(1)$ theory and show that the slice independence of the refined 
closed topological string amplitude gives the combinatorial identity \eqref{5Dconjecture}. 
It turns out that the validity of \eqref{5Dconjecture} is equivalent to the agreement of 
two known ways of computing the $\chi_y$ genus of the Hilbert scheme of points on ${\mathbb C}^2$. 
In section 3 we review the relation of the homological invariants of the colored
Hopf link and the refined open topological string amplitudes following
\cite{GIKV}. We also introduce the conjecture presented in \cite{Taki2} and
explain how it is related to the problem of changing the preferred direction.
Section 4 is the main part of the paper. We prove the formula
\eqref{main} by making use of the Macdonald operator
on the space of the symmetric functions. In section 5 by using the formula
we provide a general formula for the superpolynomial 
of the Hopf link colored with any pair $(\la,\mu)=(1^r,1^s)$ of totally
anti-symmetric representations. We also show that homological invariants obtained from 
\eqref{GIKVsum} are actually two variable Laurent polynomials
in $\mbq$ and $\mbt$ with positive integer coefficients.
Some examples of the superpolynomial for representations other than the totally anti-symmetric
ones are worked out in section 6. By the base change between the Schur functions
and the Macdonald functions, we can replace one of the anti-symmetric representations 
$(1^r,1^s)$ with any representation $\la$. 
We compute the superpotential explicitly when $\la$ is the symmetric representation 
or the hook representation. 
We also argue some general structure of the superpolynomial of the Hopf link
colored with $(\la,1^s)$.
In appendix combinatorial identities coming from the slice invariance 
of the toric diagram of $U(1)^N$ theory are presented.

The following notations are used through this article.
$q$ and  $t$ are formal parameters otherwise stated.
Let $\lambda$ be a Young diagram,
{\it i.e.}, 
a partition $\lambda = (\lambda_1,\lambda_2,\cdots)$,
which is a sequence of non-negative integers such that
$\lambda_{i} \geq \lambda_{i+1}$ and 
$|\lambda| = \sum_i \lambda_i < \infty$.
$\lambda^\vee $ denotes its conjugate (dual) diagram.
$\ell(\lambda) = \lambda^\vee_1$ is the length and 
$|\lambda| = \sum_i \lambda_i$ is the weight.
We also use $|| \lambda ||^2 = \sum_i \lambda_i^2$
and $\kappa(\lambda) := || \lambda ||^2 -  || \lambda^\vee ||^2$.
For each square $s=(i,j)$ in $\lambda$, 
$a(s):= \lambda_i-j$ 
and 
$\ell(s):= \lambda^\vee_j-i$ 
are the arm length and  
the leg length, 
respectively\footnote{In \cite{IKV} the definitions of the arm length 
and the leg length are exchanged.}.
%
The notations for the symmetric functions in $x=(x_1,x_2,\cdots)$ are
the Macdonald function $P_\la(x;q,t)$, the Schur function $s_\la(x)$ and 
the elementary symmetric function $e_\la(x)$, respectively.
The algebra of symmetric functions $\Lambda$ 
can be identified as a polynomial ring in the
power sums $\{p_n(x)\}_{n=1}^\infty$ with $p_n(x) = \sum_{i=1}^\infty x_i^n$. 
All symmetric functions in this article are treated as 
polynomials in the power sum symmetric functions $(p_1,p_2,\cdots)$.
For $|t^{-1}|<1$, 
$x=q^\lambda t^\rho$ means $x_i = q^{\lambda_i} t^{\ha-i}$.
By definition a specialization of the symmetric functions is
an algebra homomorphism $\rho : \Lambda \to R$. 
For example one may take $R={\mathbb C}$.
Since we treat the symmetric functions as polynomials in $p_n$,
we can define a specialization by giving the images $\rho(p_n)$
of the power sums under $\rho$. For example the specialization 
of the symmetric functions with $x = q^{\lambda} t^{\rho}$ is defined by
\be
p_n(q^\lambda t^\rho) 
:=
\sum_{i=1}^N (q^{n\lambda_i}-1)t^{n(\ha-i)}
+ {1 \over t^{n\over 2} - t^{-{n\over 2}} }
=
\sum_{i=1}^N q^{n\lambda_i}t^{n(\ha-i)}
+ {t^{-nN} \over t^{n\over 2} - t^{-{n\over 2}} },
\ee
which is independent of $N$ as long as  $N\geq \ell(\la)$. 
In this specialization the algebra $R={\mathbb Q}(q,t)$ is the field of rational functions of $q$ and $t$. 
We also note that for $p_n(x,y) := p_n(x) + p_n(y)$,
\be
p_n(c q^\lambda t^\rho, L t^{-\rho} ) 
=
c^n\sum_{i=1}^{\ell(\la)} (q^{n\lambda_i}-1)t^{n(\ha-i)}
+ {c^n - L^n \over t^{n\over 2} - t^{-{n\over 2}} },
\qquad 
c,L \in\bC. 
\ee
An involution $\iota$ acting on the power sum function $p_n(x)$ is defined as
$p_n(\iota x) := - p_n(x)$.

The definition of the $q$-integer in this article\footnote{
We follow the standard convention in knot theory.
The definition $\ds{[n]_q := q^{n/2} - q^{-n/2}}$
is also used in the literatures.} is 
$\ds{[n]_q := \frac{q^{n} - q^{-n}}{q - q^{-1}}
= q^{1-n}\frac{1-q^{2n}}{1-q^2}}$.
In terms of $\ds{[n]!_q = [n]_q [n-1]_q \cdots [1]_q}$,
the $q$-binomial coefficient is 
\beq
\left[
{n\atop k}
\right]_q
:=\frac{[n]!_q}{[n-k]!_q[k]!_q}
= q^{k(n-k)}
\prod_{i=1}^k
\frac{1-q^{2(n-k+i)}} {1-q^{2i}}~.
\eeq
For $n,k\in\bN$ both the $q$-integer and the $q$-binomial coefficient are
polynomials in $q$ with positive integer coefficients. 
Finally, we often use $\tQ := vQ$ with
$v :=\left({q/t}\right)^{\ha}$.

\Section{Five-dimensional $U(1)$ theory with extra matter}


Changing the preferred direction in the diagrammatic
computation of the refined topological vertex often gives a highly non-trivial combinatorial
equality that involves a summation over partitions. 
In this section we provide an example which shows a curious
connection of the change of the preferred direction and a combinatorial identity 
in enumerative geometry. 


By using the Macdonald function $P_\la(x;q,t)$,
the refined topological vertices are defined as \cite{AK2}
\ba
\Ciio\mu \lambda  \nu{q}{t}
&:=&
P_\lambda  (t^\rho;q,t)
\sum_\sigma
P_{\mu^\vee /\sigma^\vee}(-\iota t^{\lambda^\vee}q^{\rho};t,q) \ 
P_{\nu /\sigma}(q^{\lambda } t^{\rho};q,t)
v^{|\sigma|-|\nu|}
\fla\nu qt ^{-1},
\cr
\Cooi\mu \lambda\nu{q}{t}
&:=&
\Ciio{\mu^\vee}{\lambda^\vee}{\nu^\vee}{t}{q}
(-1)^{|\lambda|+|\mu|+|\nu|}
\\
&=&
P_{\lambda^\vee}(-q^{\rho};t,q)
\sum_\sigma
P_{\nu^\vee/\sigma^\vee}(-t^{\lambda^\vee} q^{\rho};t,q) \
P_{\mu  /\sigma}(\iota q^{\lambda  } t^{\rho};q,t)
v^{-|\sigma|+|\nu|}
\fla\nu qt ,
\cr
\Coii\mu \lambda  \nu{q}{t}
&:=&
\Ciio\mu\lambda\nu{q}{t}
v^{|\mu|+|\nu|}
\fla\mu qt \fla\nu qt ,
\cr
\Cioo\mu\lambda \nu{q}{t}
&:=&
\Cooi\mu \lambda \nu{q}{t}
v^{-|\mu|-|\nu|}
\fla\mu qt ^{-1}\fla\nu qt ^{-1},
\ea
with the framing factor
\be
f_\lambda(q,t)
:=
\prod_{(i,j)\in\lambda}(-1)
q^{\lambda_i-j+\ha}t^{-\lambda^\vee_j+i-\ha}.
\label{eq:framing}
\ee
The lower and the upper indices correspond to 
the incoming and the outgoing representations,
respectively,
and the edges of the topological vertex are ordered clockwise.
The middle index $\lambda$ is the representation for the preferred direction.


\FigUone

As has been discussed in \cite{IKS, AK2}, 
the refined topological string amplitude for the diagrams in
Fig.\ 1 gives the generating function of the equivariant $\chi_y$ 
genus  of the Hilbert scheme $\mathrm{Hilb}^n ({\mathbb C}^2)$
of $n$ points on ${\mathbb C}^2$.
These diagrams are one-loop diagrams where we identify two external vertical edges
and arise from the geometric engineering of five-dimensional $U(1)$ gauge theory 
with adjoint matter \cite{HIV}. In the left diagram the preferred direction is along the internal line,
while it is along the external lines in the right diagram.

The gluing rule of the refined topological vertex gives the amplitude for the left diagram
\ba
Z_{L}&:=&
\sum_{\lambda,\nu } 
Q^{|\nu |} \Lambda^{|\lambda|} 
\Ciio\bullet\lambda\nu{q}{t} 
\Cooi\bullet\lambda\nu{q}{t} \cr
&=& \sum_{\lambda, \nu } 
Q^{|\nu |}
\Lambda^{|\lambda|} 
P_\lambda(t^{\rho};q,t)
P_{\lambda^\vee} (-q^{\rho};t,q)
P_\nu (q^{\lambda}t^{\rho};q,t)
P_{\nu^\vee} (-t^{\lambda^\vee}q^{\rho};t,q).
\ea
Here $\bullet$ stands for the trivial representation, i.e.,
the zero Young diagram $(0,0,\cdots)$.
The parameters $Q$ and $\Lambda$ are related to the coupling constant 
$\tau$ of the $U(1)$ gauge theory and the mass $m$ of the adjoint 
hypermultiplet as follows;
\be
Q = e^{-m}, \qquad \Lambda \cdot Q = e^{2\pi i \tau}. 
\ee
On the other hand the right diagram gives us  the following partition function;
\ba
Z_{R} 
&:=&
\sum_{\mu,\nu}
Q^{|\nu|}\Lambda^{|\lambda|}
\Coii\nu\bullet\lambda qt
\Cioo\nu\bullet\lambda qt
\cr
&=&
\sum_{\mu,\nu, \sigma_1, \sigma_2}
Q^{|\nu|}\Lambda^{|\lambda|}
P_{\nu^\vee/\Yav\sigma 1}(-\iota q^\rho;t,q)
P_{\lambda     /\Ya \sigma 1}(       t^\rho;t,q)
P_{\lambda^\vee/\Yav\sigma 2}(-      q^\rho;t,q)
P_{\nu     /\Ya \sigma 2}( \iota t^\rho;t,q)
v^{|\Ya\sigma 1|-|\Ya\sigma 2|}.
\cr &&
\ea


The computation of $Z_{L}$ is made by 
the Cauchy formula for the Macdonald function
\be
\sum_\lambda
P_{\lambda/\mu} (x;q,t) P_{\lambda^\vee/\nu^\vee} (y;t,q)
=
\Pi_0(x,y)
\sum_\eta
P_{\mu^\vee/\eta^\vee} (y;t,q) P_{\nu/\eta} (x;q,t),
\label{eq:skewCauchy}
\ee
where
\be
\Pi_0(-x,y)
:= 
\Exp{
-\sum_{n>0}
{1
\over n} p_n(x) p_n(y)
}
=
\prod_{i,j}(1- x_i y_j).
\label{eq:Pizero}
\ee
Note that for $c\in\bC$, $\Pi_0(cx,y) = \Pi_0(x,cy)$
and for the involution $\iota(p_n)=-p_n$,
$\Pi_0(\iota x,y) = \Pi_0(x,\iota y) = \Pi_0(x,y)^{-1}$.
We also use the following adding formula
\be
\sum_\mu
P_{\lambda/\mu} (x;q,t) 
P_{\mu/\nu} (y;q,t) 
= 
P_{\lambda/\nu} (x,y;q,t).
\label{eq:addSkewMacdonald}
\ee
From the formula of the principal specialization
\be
P_\lambda(t^{\rho};q,t)
=
\prod_{s\in\lambda} 
{
(-1) t^{\ha} q^{a(s)}
\over 
1-q^{a(s)} t^{\ell(s)+1} 
},
\qquad
P_{\lambda^\vee}(-q^{\rho};t,q)
=
\prod_{s\in\lambda} 
{
(-1)q^{-\ha} q^{-a(s)}
\over 
1-q^{-a(s)-1} t^{-\ell(s)} 
},
\label{eq:LargeNPrincipalSpecialization}
\ee
and the Cauchy formula
\eqref{eq:skewCauchy} for $\mu=\nu=\bullet$,
we have
\be
Z_{L}
= \sum_{\lambda} 
\Pi_0(-Q q^{\lambda}t^{\rho},\, t^{\lambda^\vee}q^{\rho})
\prod_{s\in\lambda}
{v^{-1}\Lambda
\over 
(1-q^{ a(s)  } t^{ \ell(s)+1})
(1-q^{-a(s)-1} t^{-\ell(s)  })}.
\ee
If we define the perturbative part by
$
\Zm{pert}{L}
:=
\sum_{\nu } 
Q^{|\nu |} 
\Ciio\bullet\bullet\nu{q}{t} 
\Cooi\bullet\bullet\nu{q}{t}
=
\Pi_0(-Q t^{\rho},\, q^{\rho})
$,
which is independent of $\Lambda$, 
then the instanton part
$
\Zm{inst}{L}
:= Z_{L}/\Zm{pert}{L}
$
is
\ba
\Zm{inst}{L}
&=&
\sum_{\lambda} 
\prod_{s\in\lambda}
v^{-1}\Lambda
{
1- \tQ q^{ a(s)  } t^{ \ell(s)+1}\over 
1-  q^{ a(s)  } t^{ \ell(s)+1}
}
{
1- \tQ q^{-a(s)-1} t^{-\ell(s)  }\over 
1-         q^{-a(s)-1} t^{-\ell(s)  }
},
\label{eq:chiy}
\ea
where $\tQ = v Q$. 
Note that  we  have used
 ((2.10) in \cite{AK2})
\ba
N_{\lambda,\mu}(Q;q,t)  
&:=&
\prod_{(i,j)\in\lambda} 
\left( 1 - Q\, q^{\lambda_i-j} t^{\mu^\vee_j-i+1} \right)
\prod_{(i,j)\in\mu } 
\left( 1 - Q\, q^{-\mu_i+j-1} t^{-\lambda^\vee_j+i  } \right)
\cr
&=&
\Pi_0(-v^{-1} Q\, q^{\lambda}t^{\rho},\  t^{\mu^\vee}q^{\rho})
/
\Pi_0(-v^{-1} Q\, t^{\rho},\  q^{\rho}).
\label{eq:NekIIp}
\ea


The computation of $Z_{R}$ is more involved and we have to employ 
the trace formula%
\footnote{
This is obtained by applying the automorphism 
$
\omega_{q,t}(p_n) 
:=
(-1)^{n-1}{1-q^n \over 1-t^n} p_n
$
on $y$ and $w$
in (B.26) of \cite{AK1}.
}%
 (see (B.26) in \cite{AK1}),
which is obtained by successively using \eqref{eq:skewCauchy},
\ba
&&\hskip -10pt
\sum_{\{\lambda, \eta, \mu, \sigma\}}
P_{\mu^\vee    /\sigma^\vee} (x;t,q) 
P_{\lambda     /\sigma     } (y;q,t) 
P_{\lambda^\vee/\eta^\vee  } (z;t,q) 
P_{\mu         /\eta       } (w;q,t) 
\alpha^{|\lambda|}
\beta^{-|\eta|}
\gamma^{|\mu|}
\delta^{-|\sigma|}
\cr
&=&
\prod_{k\geq 0} {1\over 1-c^{k+1}}
\Pi_0(y, \  \alpha c^k  z)
\Pi_0(y, \  \delta^{-1} c^{k+1} x )
\Pi_0(w, \  \g c^k  x)
\Pi_0(w, \  \b^{-1} c^{k+1} z )
\cr
&=&
\exp\left\{-\sum_{n>0}{(-1)^{n}\over n}
{1\over 1-c^n} 
\left(
\a         ^n p_n(y) p_n(z)
+{c^n\over \delta^n} p_n(y) p_n(x)
\right.\right.
\cr
&&\hskip 125pt
\left.\left.
+\g         ^n p_n(w) p_n(x)
+{c^n\over \beta^n} p_n(w) p_n(z)
-c^n
\right)\right\},
\label{eq:TraceFormulaII}
\ea
with
$c:=\a \g /\b \delta $.
Using this trace formula, 
we can compute the partition function $Z_{R}$ as follows;
if we put
$(\a ,\b ,\g ,\delta )=(\Lambda,v^{-1},Q,v)$ and
$(x,y,z,w)=(-\iota q^\rho,t^\rho,-q^\rho,\iota t^\rho)$,
then $c=Q\Lambda$. 
Hence from \eqref{eq:TraceFormulaII}
\ba
Z_{R}
&=&
\prod_{k\geq 0} 
{
\Pi_0(t^\rho,-\Lambda c^k q^\rho)
\Pi_0(t^\rho,-Q      c^k q^\rho)
\over
\Pi_0(t^\rho,-v c^{k+1}q^\rho)
\Pi_0(t^\rho,-v^{-1}c^{k+1}q^\rho)
}
{1\over 1-c^{k+1}}
\cr
&=&
\exp\left\{
-\sum_{n>0} {1\over n} {1\over 1-c^n}
\left(
{
(\Lambda^n + Q^n) - (v^n + v^{-n})c^n  
\over 
(t^{n\over 2} - t^{-{n\over 2}})
(q^{n\over 2} - q^{-{n\over 2}})
}
-c^n
\right)\right\}.
\ea
As before we define the perturbative part by 
\be
Z_R^{\rm pert}
:=
Z_R(\Lambda = 0)
=
\exp\left\{
-\sum_{n>0}
\frac{Q^n}
{n
(t^{n\over 2} - t^{-{n\over 2}})
(q^{n\over 2} - q^{-{n\over 2}})
}
\right\}.
\ee
Then the instanton part
$
Z_{R}^{\rm inst}
:=
Z_{R}/Z_{R}^{\rm pert}
$
is
\be
Z_{R}^{\rm inst}
=
\exp\left\{
-\sum_{n>0} {1\over n} {\Lambda^n\over 1-c^n}
{
(Q^n - u^n)(Q^n - u^{-n})
\over 
(t^{n\over 2} - t^{-{n\over 2}})
(q^{n\over 2} - q^{-{n\over 2}})
}
\right\},
\label{eq:LoopChiy}
\ee
with $u:=(qt)^\ha$.
If the topological partition function is independent of the choice of
the preferred direction in our one-loop diagram, $Z_{L}^{\rm inst}=Z_{R}^{\rm inst}$ and
we should have
\ba
&&\hskip-60pt
\sum_{\lambda} 
\Lambda^{|\lambda|}
\prod_{s\in\lambda}
{
1-y q^{ a(s)  } t^{ \ell(s)+1}\over 
1-  q^{ a(s)  } t^{ \ell(s)+1}
}
{
1-y q^{-a(s)-1} t^{-\ell(s)  }\over 
1-         q^{-a(s)-1} t^{-\ell(s)  }
}
\cr
&=&
\exp\left\{
\sum_{n>0} {1\over n} {\Lambda^n\over 1-\Lambda^n y^n}
{
(1-t^n y^n)(1-q^{-n}y^n)
\over 
(1-t^n)(1-q^{-n})
}
\right\},
\label{eq:conjecture}
\ea
where we have introduced $y := v Q$ for convenience and 
rescaled $\Lambda$ by $v^{-1} = (t/q)^{\frac{1}{2}}$.

Thus the slice invariance requires the highly non-trivial identity \eqref{eq:conjecture}. 
One can check the validity of \eqref{eq:conjecture} 
for  special cases $y=0$ and $y=1$. When $y=0$ it reduces to 
\be
\sum_{\lambda} 
\prod_{s\in\lambda}
{\Lambda
\over 
(1-  q^{ a(s)  } t^{ \ell(s)+1})
(1-q^{-a(s)-1} t^{-\ell(s)  })
}
=
\exp\left\{
\sum_{n>0} {\Lambda^n\over n}
{1
\over 
(1-t^n)(1-q^{-n})
}
\right\},
\ee
which was proved by Nakajima and Yoshioka \cite{NY}. 
The proof in \cite{NY} is geometric. 
Namely we can see that
the right hand side is the generating function of the Hilbert series of
$\mathrm{Hilb}^n ({\mathbb C}^2)$ as follows;
\beq
\sum_{n=0}^\infty \Lambda^n
\mathrm{ch}~H^0(\mathrm{Hilb}^n ({\mathbb C}^2), {\cal O})
= \prod_{k, \ell \geq 0} {1 \over (1- t^{k} q^{-\ell} \Lambda)} 
=\exp\left\{
\sum_{n>0} {\Lambda^n\over n}
{1\over (1- t^{n})(1- q^{-n})}
\right\}.
\eeq
On the other hand the left hand side arises from a computation of the generating function
by the localization theorem for toric action, where the fixed points of the toric action are
labeled by partitions.
We also have a combinatorial proof based on
the Cauchy formula
\eqref{eq:skewCauchy} for $\mu=\nu=\bullet$.
The formula of the principal specialization 
\eqref{eq:LargeNPrincipalSpecialization}
implies the desired identity.
%
%
For $y=1$ the conjecture is simply
\be
\sum_{\lambda} 
\Lambda^{|\lambda|}
=
\exp\left\{
\sum_{n>0} {1\over n} {\Lambda^n\over 1-\Lambda^n}
\right\}
= \prod_{n=1}^\infty (1 - \Lambda^n)^{-1},
\ee
which is nothing but the generating function of the number of partitions. 
Physically $y= e^{-m}$ in terms of the mass $m$ of the adjoint matter. 
Thus the equality \eqref{eq:conjecture} interpolates the massless theory (${\mathcal N}=4$ theory)
and the infinitely massive theory  (${\mathcal N}=2$ theory).

As has been pointed out in \cite{LLZ}, 
the right hand side of \eqref{eq:conjecture}  is identified as the generating function of 
the orbifold equivariant $\chi_y$ genera of the symmetric product 
$\mathrm{Sym}^n({\mathbb C}^2) := ({\mathbb C}^2)^n/S_n$. 
The Hirzebruch $\chi_y$ genus of a manifold $M$ is defined by
\beq
\chi_{y} (M) := \sum_{p,q} (-1)^q y^p \dim H^q (M, \wedge^p T^*M)~, \label{chiy}
\eeq
and we will consider the equivariant version of \eqref{chiy} where we use
the equivariant cohomology $H^q_G (M, \wedge^p T^*M)$.
Since 
\beqa
H^0_{T^2} ({\mathbb C}^2, \Lambda^p T^*{\mathbb C}^2) 
&=& 
\begin{cases}
{\mathbb C}[z_1, z_2], \quad (p=0) \\
{\mathbb C}[z_1, z_2] dz_1\oplus {\mathbb C}[z_1, z_2] dz_2, \quad (p=1) \\
{\mathbb C}[z_1, z_2] dz_1 \wedge dz_2, \quad (p=2) \\
\end{cases}  \\
H^q_{T^2} ({\mathbb C}^2, \Lambda^p T^*{\mathbb C}^2) 
&=& 0, \qquad (q \neq 0) 
\eeqa
the generating function of the equivariant $\chi_y$ genus of
${\mathbb C}^2$ is given by
\be
\chi_{-y}({\mathbb C}^2) = \sum_{k, \ell=0}^\infty t_1^k t_2^\ell  \left(1 
-y (t_1+ t_2) + y^2 t_1t_2\right)
= \frac{(1-y t_1)(1-y t_2)}{(1-t_1)(1-t_2)} ~.
\ee
A formula of the elliptic genera of the symmetric product $\mathrm{Sym}^n M$ was
proposed in \cite{DMVV}, which gives the elliptic genera of 
$\mathrm{Sym}^n M$ in terms of those of the underlying K\"ahler manifold $M$,
see also \cite{BL}. 
The formula tells us that
\beq
\sum_{n=0}^\infty z^n \mathit{Ell}_{\mathrm{orb}}(\mathrm{Sym}^n M; y,q)
= \prod_{i=1}^\infty \prod_{\ell, m} \frac{1}{(1- z^i y^\ell q^m)^{c(mi,\ell)}}~, \label{DMVV}
\eeq
where $c(m,\ell)$ are the coefficients of the elliptic genus of $M$;
\beq
\mathit{Ell}(M; y,q) = \sum_{m, \ell} c(m,\ell) y^\ell q^m~.
\eeq
Since the $\chi_y$ genus is obtained from the elliptic genus
by $\mathit{Ell}(M;y,q=0) = y^{-\frac{\dim M}{2}} \chi_{-y}(M)$,
from DMVV formula \eqref{DMVV} we find
\be
\sum_{n \geq 0} \Lambda^n \chi_{-y}(\mathrm{Sym}^n({\mathbb C}^2)) = 
\prod_{m \geq 1} \prod_{k, \ell \geq 0} 
\frac{\left(1- \Lambda^m y^m  t_1^{k+1} t_2^\ell \right)
\left(1- \Lambda^m y^m  t_1^k t_2^{\ell+1}\right)}
{\left(1- \Lambda^m y^{m-1} t_1^k t_2^\ell\right)
\left(1- \Lambda^m y^{m+1} t_1^{k+1} t_2^{\ell+1} \right)}~, \label{sym}
\ee
which means
\ba
\log\left\{ \sum_{n \geq 0} \Lambda^n \chi_{-y}(\mathrm{Sym}^n({\mathbb C}^2)) \right\}
&=& 
\sum_{n \geq 1} \sum_{m \geq 1} \sum _{k, \ell \geq 0} 
\frac{1}{n} \Lambda^{nm} y^{n(m-1)} t_1^{nk} t_2^{n\ell} (1- y^n t_1^n)(1- y^n t_2^n) \cr
&=& \sum_{n \geq 1} \frac{1}{n} \frac{\Lambda^n}{1- \Lambda^n y^n}
\frac{(1- y^n t_1^n)(1- y^n t_2^n) }{(1 - t_1^n)(1- t_2^n)}~.
\ea
On the other hand the localization theorem tells us that  the left hand side is the equivariant 
$\chi_y$ genera of the Hilbert scheme $\mathrm{Hilb}^n ({\mathbb C}^2)$ of points on ${\mathbb C}^2$. 
Thus the slice invariance \eqref{eq:conjecture} geometrically means
that the equivariant $\chi_y$ genera of $\mathrm{Hilb}^n ({\mathbb C}^2)$ which gives
a resolution of the symmetric product agrees with
the orbifold equivariant $\chi_y$ genera of $\mathrm{Sym}^n ({\mathbb C}^2)$. 
There are several evidences that they agree in mathematically rigorous sense. 
For equivariant Euler character the agreement was proved in \cite{NY}. 
Furthermore, in \cite{Wae} it was proved that the equivariant version of \eqref{DMVV} is valid for
the Hilbert scheme $\mathrm{Hilb}^n ({\mathbb C}^2)$, which implies that \eqref{sym} holds
not only for $\mathrm{Sym}^n ({\mathbb C}^2)$ but also for $\mathrm{Hilb}^n ({\mathbb C}^2)$. 
Finally the following formula of the generating function of 
the $\chi_y$ genera of the Hilbert scheme of a smooth projective surface $S$ was
established in \cite{EGL};
\beq
\sum_{n=0}^\infty  z^n \chi_{-y} (\mathrm{Hilb}^n (S))
= \Exp{\sum_{m=1}^\infty \frac{\chi_{-y^m}(S)}{1- y^m z^m} \frac{z^m}{m} }~.
\eeq
This formula for the equivariant case would give us a proof of  \eqref{eq:conjecture}.


\Section{Refined topological vertex and homological link invariants}


%


In \cite{GSV} it was argued that homological link invariants are
related to a refinement of the BPS state counting in topological open string theory. 
Based on this proposal, in \cite{GIKV} a conjecture on homological link invariants of the Hopf link 
from the refined topological vertex has been provided. Let us review their proposal briefly. 
Let $L$ be an oriented link in $S^3$ with $\ell$ components. We consider 
homological invariants of $L$ whose components are colored by representations
$R_1, \cdots, R_\ell$ of the Lie algebra $\mathfrak{sl}(N)$. This means that we have
a doubly graded homology theory ${\cal H}_{i,j}^{\mathfrak{sl}(N); R_1, \cdots, R_\ell} (L)$ 
whose graded Poincar\'e polynomial is
\beq
\overline{\cal P}_{\mathfrak{sl}(N); R_1, \cdots, R_\ell} (\mbq,\mbt) = \sum_{i,j \in {\mathbb Z}} 
\mbq^i \mbt^j \dim {\cal H}_{i,j}^{\mathfrak{sl}(N); R_1, \cdots, R_\ell} (L)~. \label{super1}
\eeq
Substitution of $\mbt=-1$ gives the unnormalized link polynomial 
$\overline{P}_{\mathfrak{sl}(N); R_1, \cdots, R_\ell} (\mbq) 
= \overline{\cal P}_{\mathfrak{sl}(N); R_1, \cdots, R_\ell} (\mbq,-1)$.
The normalized invariants ${P}_{\mathfrak{sl}(N); R_1, \cdots, R_\ell} (\mbq)$ are obtained by
dividing by the invariants of the unknot. However, in the following we only consider 
the unnormalized one.

The physical interpretation of homological invariants as the BPS state counting leads to
a prediction on the dependence of the link homologies on the rank $N-1$. 
It has been conjectured \cite{DGR, GW} that there exists a \lq\lq superpolynomial\rq\rq\ 
$\overline{\cal P}_{R_1, \cdots, R_\ell} (\mba, \mbq,\mbt)$ which is a rational
function in three variables such that
\beq
\overline{\cal P}_{\mathfrak{sl}(N); R_1, \cdots, R_\ell} (\mbq,\mbt) =
\overline{\cal P}_{R_1, \cdots, R_\ell} (\mba=\mbq^N, \mbq,\mbt)~. \label{super2}
\eeq
We can reproduce the polynomial invariants of the colored Hopf link from
topological open string amplitudes on the conifold with appropriate brane
 configuration \cite{Wit, GV, OV, LMV, MV}.  Since they are
computed by the method of topological vertex, it is natural to expect that
a refined version of the topological vertex gives a homological
version of the Hopf link invariants. 
Based on the refined topological vertex introduced in \cite{IKV}, 
the superpolynomial for  the colored Hopf link was proposed in \cite{GIKV};
\beq
\overline{\cal P}_{\lambda, \mu}( \mba, \mbq, \mbt)
= (-1)^{|\lambda| + |\mu|}
 \left( Q^{-1} \sqrt{\frac{q_1}{q_2}} \right)^{\frac{1}{2}(|\lambda| + |\mu|)}
 \left(\frac{q_1}{q_2} \right)^{|\lambda||\mu|}
  \frac{Z_{\lambda,\mu}(q_1, q_2, Q)}{Z_{\bullet,\bullet}(q_1, q_2, Q)}~,
\label{GIKVconj2}
\eeq
where
\beq
Z_{\lambda,\mu}(q_1,q_2,Q) =
\sum_{\eta} (-Q)^{|\eta|} q_2^{\frac{1}{2}||\eta^\vee||^2} q_1^{\frac{1}{2}||\eta||^2}
\tilde Z_{\eta^\vee}(q_1,q_2) \tilde Z_{\eta}(q_2,q_1)
s_\lambda(q_1^{-\eta} q_2^{-\rho}) s_\mu(q_1^{-\eta} q_2^{-\rho})~. \label{HLZ}
\eeq
The natural variables of symmetric functions $(q_1, q_2)$ are related to the
variables of the superpolynomial by
\beq
\sqrt{q_2} = \mbq, \quad \sqrt{q_1} = - \mbt\mbq, \quad Q = - \mbt\mba^{-2}.
\eeq
The factor $\tilde Z_\eta$ in \eqref{HLZ} is given by\footnote{The definitions of 
the arm length $a(s)$ and the leg length $\ell(s)$ in \cite{GIKV} are
exchanged, compared with the standard ones, for example in \cite{Mac}.}
\beq
\tilde Z_\eta (q_2, q_1) = \prod_{(i,j) \in \eta} \left(1 - q_2^{\eta_j^\vee -i +1} q_1^{\eta_i -j} \right)^{-1}
= \prod_{s \in \eta} \left(1 - q_2^{\ell(s)+1} q_1^{a(s)} \right)^{-1}~,
\eeq
and, therefore, related to the following specialization of the Macdonald function;
\beq
P_\lambda(t^\rho; q,t) = t^{-\frac{1}{2}||\lambda^\vee||^2} \prod_{s \in \lambda}
\left(1 - q^{-a(s)} t^{-\ell(s)-1} \right)^{-1}~.
\eeq
Namely if we identify $(q_1, q_2) \equiv (q^{-1}, t^{-1})$, then we find
\beq
Z_{\lambda,\mu}(q_1,q_2,Q) = \sum_\eta (-Q)^{|\eta|} P_{\eta^\vee}(q^\rho; t,q) 
P_\eta(t^\rho; q,t) s_\lambda(q^\eta t^\rho) s_{\mu}(q^\eta t^\rho)~. \label{partitionfn}
\eeq
Note that this expression is manifestly symmetric in $\lambda$ and $\mu$.
But due to the summation over $\eta$, it is not clear at all that $Z_{\lambda,\mu}$ is
a polynomial in $Q$. Furthermore, since the superpolynomial $\overline{\cal P}_{\lambda, \mu}$ is
defined by \eqref{super1} and \eqref{super2}, we would like to check that, when $\mba=\mbq^N$,  
the superpolynomials derived from \eqref{partitionfn} are in fact Laurent polynomials
in $\mbq$ and $\mbt$ with positive integer coefficients. Thus it is desirable to
have a closed formula of $Z_{\lambda,\mu}$ without a summation over partitions.

\FigSlicingFlop

The configuration that leads to the proposal \eqref{GIKVconj2} is 
the toric diagram of the resolved conifold with two Lagrangian brane
insertions (see Fig.\ 2). Two branes are inserted on the same side of 
external edges. But they are on the different vertices. 
In \cite{GIKV} the refined topological string amplitude for
this diagram was computed by choosing
the internal line as the preferred direction.
This was the reason why the summation over the partition $\eta$, which
was attached to the internal line, remained in the proposal \eqref{GIKVconj2}
for the superpolynomial. As has been pointed out by Taki \cite{Taki2},
if we assume the slice invariance that means
the partition function computed by the refined topological vertex is 
independent of the choice of the preferred direction, 
we may have a formula of $\overline{\cal P}_{\lambda, \mu}$ in a closed
form without a summation over partitions.
Namely we first change the preferred direction from the
internal line to the external lines as shown in Fig.\ 2. 
Then we make a flop operation to move two Lagrangian branes
to the same vertex. 
When we compute the topological string amplitude 
based on the final diagram, we can perform the summation
over the partitions of the internal line and obtain
\begin{align}
\overline{\cal P}_{\lambda, \mu}( \mba, \mbq, \mbt)
&= \left(Q^{-1} \sqrt{\frac{q_1}{q_2}} \right)^{-\frac{1}{2}(|\lambda| + |\mu|)}
 \left(\frac{q_1}{q_2}\right)^{-n(\mu^\vee)+|\lambda||\mu|}
q_2^{-\frac{1}{2}(\kappa(\lambda)+\kappa(\mu))}  \CR
&~~ s_\mu(q_2^{-\rho}) s_\lambda(q_1^{-\mu} q_2^{-\rho}, Q^{-1}\sqrt{\frac{q_1}{q_2}} q_2^\rho)
\prod_{(i,j) \in \mu} \left( 1- Q^{-1} q_2^{-i+ \frac{1}{2}} q_1^{\mu_i -j +\frac{1}{2}} \right)~.
\label{Takiconj}
\end{align}
Since the flop operation in the computation of the refined topological vertex was
worked out completely in \cite{AK2, Taki2}, we only have to assume the 
slice independence to derive \eqref{Takiconj}. In other words the validity of 
\eqref{Takiconj} is equivalent to the slice invariance. 
In the next section we prove \eqref{Takiconj} agrees to the original proposal 
\eqref{GIKVconj2} up to an overall normalization, if both 
$\lambda$ and  $\mu$ are totally antisymmetric representations.
The proof relies on various properties of the Macdonald functions 
and is independent of the computation of the refined topological vertex. 
This means the slice invariance holds in this case. However, in section 6
we will see that we need a few correction terms to the formula
\eqref{Takiconj} for general representations. 


\Section{Macdonald operator and summation over partitions}


In this section we will work out the summation over the partitions $\eta$ in \eqref{partitionfn}, 
when both $\lambda$ and $\mu$ are totally anti-symmetric representations.
The most fundamental summation formula on the space of symmetric functions is the Cauchy
formula 
\beq
\sum_{\lambda} \frac{1}{\langle P_\lambda \vert P_\lambda \rangle_{q,t}}
P_\lambda(x; q,t) P_\lambda(y; q,t)
= \Pi(x,y;q,t) :=
\Exp{\sum_{n=1}^\infty \frac{1}{n}\frac{1-t^n}{1-q^n} p_n(x) p_n(y) }~, 
\label{Cauchy}
\eeq
where the scalar product is given by
\beq
\langle P_\lambda \vert P_\lambda \rangle_{q,t}
:= \prod_{ s \in \lambda} \frac{1- q^{a(s)+1} t^{\ell(s)}}{1- q^{a(s)} t^{\ell(s)+1}}~.
\eeq
If we try to apply it to \eqref{partitionfn}, 
the problem is the existence of two Schur functions whose specialization depends on the
partition $\eta$. Thus our strategy is to \lq\lq remove\rq\rq\ these Schur functions from the summand. 
As we will see below this is possible for totally anti-symmetric representations.
The starting point is the observation that for totally anti-symmetric representation $\lambda=(1^r)$ 
both the Schur function $s_\lambda(x)$ and the Macdonald function $P_\lambda(x; q,t)$ coincide with 
the elementary symmetric function $e_r(x)$. 
Thus we can use a trick of replacing one of the Schur functions with the Macdonald function 
to make use of the following remarkable symmetry
of the specialization of the Macdonald functions \cite{Mac}  (Ch.\ VI.6);
\be
P_\la(t^\rho;q,t)P_\mu(q^\la t^\rho;q,t) 
= P_\mu(t^\rho;q,t)P_\la(q^\mu t^\rho;q,t).
\label{eq:PPsymm}
\ee
We have
\begin{align}
Z_{1^r, 1^s}(Q; q,t) &= \sum_\eta (-Q)^{|\eta|} P_{\eta^\vee}(q^\rho; t,q)
P_{\eta}(t^\rho; q,t) P_{(1^r)}(q^\eta t^\rho; q,t) e_s (q^\eta t^\rho) \CR
&= e_r(t^\rho) \sum_\eta (-Q)^{|\eta|} P_{\eta^\vee}(q^\rho; t,q)
P_{\eta}(q^{(1^r)} t^\rho; q,t)e_s (q^\eta t^\rho)~. \label{firststep}
\end{align}
Thus we have succeeded in getting rid of one of the Schur functions from the summand. 
To eliminate the remaining elementary symmetric function in \eqref{firststep},
we introduce the Macdonald operator on the space of symmetric functions \cite{Mac}.
The Macdonald functions are characterized by the property
that they are simultaneous eigen-functions of the Macdonald operator. 
The Macdonald operator in $N$ variables $x=(x_1,\cdots,x_N)$ is given by
\be
D_N^1 := \sum_{i=1}^N \prod_{j(\neq i)}
\frac{tx_i-x_j}{x_i-x_j}
{\Diff i},
\ee
where
${\Diff{ }}$ is the $q$-shift operator defined by
${\Diff{ }} f(x) = f(qx)$. 
It is known that the Macdonald polynomials are the eigen-functions of $D_N^1$;
\be
D_N^1 P_\la(x;q,t) = \ve_{N,\la}^1 P_\la(x;q,t),
\qquad
\ve_{N,\la}^1 := t^{N-\ha}\sum_{i=1}^N q^{\la_i} t^{\ha-i}.
\ee
More generally for non-negative integer $r$, we define
the higher Macdonald operators $D_N^r$ by
$D_N^0 := 1$ and 
\be
D_N^r := t^{r(r-1)/ 2}
\sum_{| I | = r} \prod_{{i\in I \atop j \notin I}}
\frac{tx_i-x_j}{x_i-x_j}
\prod_{i\in I}{\Diff i}, \quad (1 \leq r \leq N)
\label{eq:DNr}%
\ee
where the sum is over all $r$-element subsets $I$ of $\{1,2,\cdots,N\}$.
We set $D_N^r:=0$ for $r > N$.
Let
$D_N := \sum_{r=0}^N D_N^r \tz^r$,
then the Macdonald polynomial is the eigen-function of $D_N$
\ba
&&D_N P_\la(x;q,t) = P_\la(x;q,t) \ve_{N,\la},
\cr
&&
\ve_{N,\la} := \prod_{i=1}^N (1+\tz q^{\la_i}t^{N-i})
=: \sum_{r=0}^N \tz^r e_r(q^{\la_i}t^{N-i}).
\ea
Therefore, $D_N^r$ are simultaneously diagonalized by the Macdonald 
polynomials
\be
D_N^r P_\la(x;q,t) = P_\la(x;q,t)  e_r(q^{\la_i}t^{N-i}),
\ee
and $D_N^r$ commute with each other; $[D_N^r,D_N^s]=0$, 
on the space of the symmetric functions in $N$ variables.


Since $D_N$ is not compatible with 
the restriction of the variables defined by $x_N=0$,
we need to modify it so that we can take the limit $N\rightarrow\infty$ \cite{AKOS, AOS}. 
Let us define $H$ and $H^r$ which act on $x=(x_1,x_2,\cdots)$ by the limits
$H:=\lim_{N\rightarrow\infty} H_N$ and
$H^r:=\lim_{N\rightarrow\infty} H_N^r$,
where
\ba
H_N 
&:=&
\Exp{\sum_{n>0} \frac{1}{n} \frac{(-\tz)^n}{1-t^n}}
D_N
=:
\sum_{r\geq 0} z^r H_N^r,
\qquad
z:=\tz t^{N-\ha},
\cr
H_N^r &=&
t^{(\ha-N)r} \sum_{s=0}^{\min(r,N)} 
\prod_{i=1}^{r-s} \frac{1}{ t^i-1}
\cdot
D_N^s,
\qquad
r=0,1,2,\cdots.
\ea
Here we use 
$
\sum_{n\geq 0} \tz^n\prod_{i=1}^n (t^i-1)^{-1} =
\Exp{\sum_{n>0}(-\tz)^n(1-t^n)^{-1} /n}
$
and define a new parameter $z$ by scaling $\tz$. 
Then
\ba
E_{N,\la} 
&:=&
\Exp{\sum_{n>0} \frac{1}{n} \frac{(-\tz)^n}{1-t^n}}
\ve_{N,\la} 
\cr
&=&
\Exp{-\sum_{n>0} \frac{(-\tz)^n}{n} \( \sum_{i=1}^N q^{n\la_i} 
t^{n(N-i)}+ \frac{1}{t^n-1} \)}
\cr
&=&
\Exp{-\sum_{n>0} \frac{(-z)^n}{n} p_n(q^\la t^\rho)}
=
\sum_{r\geq 0} z^r e_r(q^\la t^\rho).
\ea
Since this is independent of $N$ as long as it is larger than the length $\ell(\la)$ of the partition, 
we have
\ba
H_N P_\la(x;q,t) &=& P_\la(x;q,t) E_{N,\la},
\cr
H_N^r P_\la(x;q,t) &=& P_\la(x;q,t) e_r(q^\la t^\rho). 
\label{eq:EigenHN}
\ea
Now by \eqref{eq:EigenHN} we can eliminate the elementary symmetric 
function in \eqref{firststep} and apply 
the Cauchy formula \eqref{Cauchy} 
as follows;
\begin{align}
Z_{1^r, 1^s}(Q; q,t) 
&= e_r (t^\rho) \sum_\eta \frac{(\tQ)^{|\eta|}} {\langle P_\eta | P_\eta \rangle_{q,t}}
P_{\eta}(t^{-\rho}; q,t) P_{\eta}(q^{(1^r)} t^\rho; q,t)e_s (q^\eta t^\rho) \CR
&= e_r (t^\rho) \sum_\eta \frac{P_{\eta}(\tQ q^{(1^r)} t^\rho; q,t) } 
{\langle P_\eta | P_\eta \rangle_{q,t}}
\left.\left( H^s(x) P_{\eta}(x; q,t) \right) \right|_{x=t^{-\rho}}\CR
&= e_r (t^\rho) \left.\left( H^s(x) \Pi(x, \tQ q^{(1^r)} t^\rho; q,t)\right)\right|_{x=t^{-\rho}}~, \label{secondstep}
\end{align}
where we have also used the following formula proved in \cite{AK1}
\be
P_{\mu^\vee}(-t^{\lambda^\vee}q^{\rho};t,q)
=
\frac{(q/t)^{\frac{|\mu|}{2}}}{\langle P_\mu \vert P_\mu \rangle_{q,t} }
P_{\mu} (q^{-\lambda}t^{-\rho};q,t).
\label{eq:Pdual}
\ee

The final step is the evaluation of $H^s(x) \Pi(x, \tQ q^{(1^r)} t^\rho; q,t)|_{x=t^{-\rho}}$.
For this purpose we have to compute the action of the shift operator ${\Diff i}$
on the Cauchy kernel $\Pi(x,y;q,t)$. 
Since ${\Diff i} p_n = \((q^n-1) x_i^n + p_n\)$ 
for the power sum $p_n = \sum_i x_i^n$,
\be
{
{\Diff i} \Pi(x,y;q,t) 
\over
\Pi(x,y;q,t) 
}
= 
\Exp{\sum_{n>0} \frac{t^n-1}{n} x_i^n p_n(y)}.
\ee
Hence,
\ba
{
\prod_{i\in I}(1-t^\ha x^i){\Diff i} \cdot \Pi(x,y;q,t) 
\over
\Pi(x,y;q,t)
}
&=& 
\Exp{\sum_{n>0} \frac{t^n-1}{n} \sum_{i\in I}x_i^n \(p_n(y)- 
\frac{t^{\frac{n}{2}}}{t^n-1}\)}
\cr
&=& 
\Exp{\sum_{n>0} \frac{t^n-1}{n} \sum_{i\in I} x_i^n 
p_n(y,t^{-\rho})}.
\label{eq:qShiftPi}%
\ea
When $x=t^{-\rho}$, since $tx_i-x_j=0$ for $j=i+1$,
only $I=\{N,N-1,\cdots,N-r+1\}$ contributes 
to the summation in \eqref{eq:DNr}.
Therefore,
\ba
D_N^r \vert_{x=t^{-\rho}}
&=&
\prod_{i=1}^r t^{i-1} \frac{1-t^{N-i+1}}{1-t^i} {\Diff{N-i+1}}~,
\cr
H_N^r \vert_{x=t^{-\rho}}
&=&
t^{(\ha-N)r}
\sum_{s=0}^r
\prod_{i=1}^{r-s} \frac{1}{t^i-1}
\prod_{j=1}^s t^{j-1} \frac{1-t^{N-j+1}}{1-t^j} {\Diff{N-j+1}}.
\ea
Then we have
\\
{\bf Proposition.}~{\it
\ba
\frac{D_N^r \Pi(x,y;q,t) \vert_{x=t^{-\rho}}}{\Pi(t^{-\rho},y;q,t)}
&=&
\Exp{\sum_{n>0} \frac{1-t^{-rn}}{n} t^{n(N+\ha)} p_n(y,t^{-\rho})}
\prod_{j=1}^r \frac{t^{j-1}}{1-t^j},
\cr
\frac{H_N^r \Pi(x,y;q,t)  \vert_{x=t^{-\rho}}}
{\Pi(t^{-\rho},y;q,t)}
&=&
(-1)^rt^{-r(r-1)/2}
e_r(y,t^{-\rho})
+ \cO(t^N),
\label{eq:HNPiPi}
\ea
with
$
\sum_{r\geq 0} (-z)^r e_r(x,y)
:=
\Exp{-\sum_{n>0}p_n(x,y) z^n/n} 
$.
}


\proof
By \eqref{eq:qShiftPi},
\ba
\frac{\prod_{i=N-r+1}^N (1-t^i){\Diff i} 
\!\cdot\!\Pi(x,y;q,t) \vert_{x=t^{-\rho}}}
{\Pi(t^{-\rho},y;q,t)}
&=& 
\Exp{\sum_{n>0} \frac{t^n-1}{n} \sum_{i=1}^r t^{n(i-\ha)} p_n(y,t^{-\rho})}
\cr
&=&
\Exp{\sum_{n>0} \frac{1-t^{-rn}}{n} t^{n(N+\ha)} p_n(y,t^{-\rho})},~~~
\ea
we get the first equation. We also have
\ba
&& \frac{H_N^r \Pi(x,y;q,t)  \vert_{x=t^{-\rho}}}
{\Pi(t^{-\rho},y;q,t)} \cr
&=&
t^{(\ha-N)r}
\sum_{s=0}^r
\Exp{\sum_{n>0} \frac{1-t^{-sn}}{n} (t^{N+\ha})^n p_n(y,t^{-\rho})}
\prod_{i=1}^{r-s} \frac{1}{t^i-1}
\prod_{j=1}^s \frac{t^{j-1}}{1-t^j}.
\ea
Then by the following lemma, we obtain the second equation.
\qed

\noindent
{\bf Lemma.}{\it
\beq
\sum_{s=0}^r \Exp{\sum_{n=1}^\infty \frac{(1- t^{-sn})}{n} t^n p_n x^n }
(-1)^s t^{\frac{1}{2}s(s-1)} \left[ r \atop s \right] 
= t^{-\frac{1}{2}r(r-1)} \prod_{k=1}^r (1-t^k) x^r e_r + O(x^{r+1})~, \label{Lemma}
\eeq
where $\sum_{r \geq 0} (-x)^r e_r = \Exp{- \sum_{n=1}^\infty p_n x^n/n}$ and
$\ds{\left[ r \atop s \right] 
:= \prod_{i=1}^s \frac{1-t^{r-s+i}}{1-t^i}}$.
}

\proof
Let us fix $k$ and consider the coefficient of $x^k$ on the left hand side.
By expanding the exponential we see that it consists of the terms with $t^{-\ell s}~(0 \leq \ell \leq k)$. 
Hence we have terms proportional to 
\beq
\sum_{s=0}^r (-t^{-\ell})^s t^{\frac{1}{2}s(s-1)} \left[ r \atop s \right],
\qquad (0 \leq \ell \leq k)~.
\eeq
However by the formula (\cite{Mac}, Chap. I-2, Example 3);
\beq
\prod_{i=0}^{r-1} (1 + t^{i} z) = \sum_{s=0}^r z^s t^{\frac{1}{2}s(s-1)} \left[ r \atop s \right],
\eeq
we see that it vanishes as long as $0 \leq \ell \leq k < r$. Finally among the coefficients of $x^r$ 
only the term with $t^{-rs}$ survives by the same reasoning as above. Hence we may compute it
with
\beq
\sum_{s=0}^r \Exp{ - \sum_{n=1}^\infty \frac{t^{-sn}}{n} (tx)^n p_n }
(-1)^s t^{\frac{1}{2}s(s-1)} \left[  r \atop s \right]
\eeq
by omitting \lq\lq$1$\rq\rq\ in the left hand side of \eqref{Lemma}.
Comparing the coefficient of $x^r$ in the relation of $p_n$ and $e_r$, 
we see the coefficient of $x^r$ is
\beq
t^r (-1)^r e_r \sum_{s=0}^r (-t^{-r})^s t^{\frac{1}{2}s(s-1)} \left[ r \atop s \right]
= (-1)^r e_r t^r \prod_{i=0}^{r-1} (1 - t^{i-r})~,
\eeq
which completes the proof. \qed

The following is the main result of this section. 

\noindent
{\bf Proposition.}~~{\it 
For $|t|<1$,
\ba
\frac{Z_{1^r, 1^s}(Q; q,t)}{Z_{\bullet, \bullet}(Q; q,t)}
&=& (-1)^s t^{-\frac{1}{2}s(s-1)} e_r(t^\rho) 
e_s(\tQ q^{(1^r)} t^\rho, t^{-\rho})
 \prod_{i=1}^r \left( 1- Q q^{\frac{1}{2}} t^{-i+\frac{1}{2}}\right)~
\cr
&=&
(-1)^{r+s} t^{-\frac{1}{2}r(r-1)-\frac{1}{2}s(s-1)} 
e_r(\tQ t^\rho, t^{-\rho}) 
e_s(\tQ q^{(1^r)} t^\rho, t^{-\rho}), \label{theorem}
\ea
where $e_r(x)$ and $e_s(x,y)$ are the elementary symmetric functions
in variables $x=(x_1, x_2, x_3, \cdots)$ and two sets of variables
$(x_1, x_2, \cdots, y_1, y_2, \cdots)$, respectively. 
}

\proof 
By taking the large $N$ limit of the proposition proved above, we see 
\begin{align}
Z_{1^r, 1^s}(Q; q,t) 
&= e_r(t^\rho) \left( H^s(x) \Pi(x, \tQ q^{(1^r)} t^\rho; q,t)\right)|_{x=t^{-\rho}} \CR
&=(-1)^s t^{-\frac{1}{2}s(s-1)} e_r(t^\rho) \Pi( t^{-\rho}, \tQ q^{(1^r)} t^\rho; q,t) 
e_s(\tQ q^{(1^r)}, t^{-\rho})~.
\end{align}
Finally the proof is completed by noting
\beq
\frac{\Pi( t^{-\rho}, \tQ q^{(1^r)} t^\rho; q,t)} {\Pi( t^{-\rho}, \tQ t^\rho; q,t)}
= N_{1^r, \bullet} (\tQ; q,t) 
= \prod_{i=1}^r \left( 1- Q q^{\frac{1}{2}} t^{-i+\frac{1}{2}} \right)~,
\eeq
where $N_{\lambda, \mu}$ is the denominator factor of Nekrasov's
partition function \eqref{eq:NekIIp},
which satisfies ((2.12) in \cite{AK2})
\be
N_{\lambda, \mu}(Q;q,t)
=
\frac{\Pi\left(Q\, q^\lambda t^\rho,q^{-\mu}t^{-\rho};q,t\right)}
{\Pi\left(Q\, t^\rho,t^{-\rho};q,t\right)}~.
\label{eq:NekIIIp}
\ee
\qed

The condition $|t|<1$ in the above proposition can be eliminated \cite{AK4}.
By the proposition it is
clear that for totally antisymmetric representations the superpolynomial 
$\overline{\cal P}_{\lambda, \mu}( \mba, \mbq, \mbt)$
is actually a polynomial in $Q$ and consequently in $\mba$,
which is not manifest in the proposal \eqref{GIKVconj2}.
We can also prove that 
it is a polynomial in $q$ and $t$ with positive integer coefficients 
when $\tQ=t^N$ with $N\in\bN$.


\Section{Homological invariants for totally anti-symmetric representations}


To accommodate the standard convention of link polynomials, we make 
the following change of variables\footnote{Note that compared with \cite{GIKV},
we have changed the translation rule by $\mbq \to \mbq^{-1}$ and 
 $\mbt \to \mbt^{-1}$. Accordingly the identification or 
 the $\mbt$-grading of $Q$ is also changed.};
\beq
\sqrt{t} = \mbq^{-1}~,\quad 
\sqrt{q} = - \mbt^{-1}\mbq^{-1}~,
\quad Q = -\mbt\mba^{-2}~,
\eeq
so that $\tQ := \left(q/t\right)^\ha Q = \mba^{-2}$. Then
\beq
\overline{\cal P}_{\lam, \mu} (\mba, \mbq, \mbt) 
= (- \mba)^{|\lambda| + |\mu|} (-\mbt)^{|\lambda||\mu|}
\frac{Z_{\lambda, \mu}(Q; q,t)}{Z_{\bullet, \bullet}(Q; q,t)}~. \label{superpol}
\eeq
Substituting the formula \eqref{theorem}, we obtain 
\beqa
\overline{\cal P}_{(1^r), (1^s)} (\mba, \mbq, \mbt) 
&=& (- \mba)^{r+s} (-\mbt)^{r\cdot s} (-1)^s \mbq^{s(s-1)}
e_r(t^\rho) e_s( \tQ q^{(1^r)}t^\rho, t^{-\rho})
\prod_{i=1}^r  \left( 1- \tQ t^{-i+1} \right), \CR
&=& \mba^{-r-s} (-\mbt)^{r\cdot s} \mbq^{s(s-1)}
e_r(t^\rho) e_s(q^{(1^r)}t^\rho, \mba^2 t^{-\rho})
\prod_{i=1}^r  \left(\mbq^{2(i-1)} - \mba^2 \right)~. \label{antisym}
\eeqa
This is a rather simple closed formula of
the superpolynomial for the Hopf link 
colored with totally anti-symmetric representations $(1^r, 1^s)$.
Note that apart from the overall factor $(-\mbt)^{r\cdot s}$, 
$\mbt$-dependence only appears in the specialization 
$x=q^{(1^r)}t^\rho$ of $e_s(x,y)$. 
By putting $\mbt = -1$ (or $q=t$) and taking the limit $\mba \to 0$
we find
\beq
\lim_{\mba \to 0} \overline{\cal P}_{(1^r), (1^s)} (\mba, \mbq, -1) 
\sim  \mba^{-r-s} \mbq^{r(r-1)+ s(s-1)} e_r(q^\rho) e_s(q^{(1^r) + \rho})~.
\eeq
We note that there is an additional factor $\mbq^{r(r-1)+ s(s-1)}$
in the normalization of \eqref{superpol},
compared with the large $N$ behavior of the polynomial invariants of the 
colored Hopf link. When $s=0$ the superpolynomial becomes independent of $\mbt$.
The formula \eqref{antisym} gives
\beq
\overline{\cal P}_{(1^r), \bullet} (\mba, \mbq, \mbt) 
= \mba^{-r} \mbq^{\frac{1}{2}r(r-1)} 
\prod_{i=1}^r \frac{\mbq^{2(i-1)} -\mba^2}{\mbq^{-i} - \mbq^i}
= \mba^{-r} \mbq^{r^2} \prod_{i=1}^r 
\frac{\mbq^{2(i-1)} -\mba^2} {1- \mbq^{2i}}~.
\eeq
This agrees with the superpolynomial of the unknot with 
a totally anti-symmetric representation in \cite{GIKV, Taki2}.
For $\mba=\mbq^N$ it reproduces the homological invariants
of the unknot, which has been proposed to give
the Hilbert series of the Grassmannian $Gr(N,r)$ \cite{GW}.

To provide more examples of our formula, 
let us look at the case where one of the representations 
is the fundamental representation. Then
the formula \eqref{antisym} is evaluated to be
\beqa
\overline{\cal P}_{(1^k),~\unitbox} (\mba, \mbq, \mbt) 
&=& \mba^{-k-1} (-\mbt)^{k} e_k(t^\rho) e_1 (q^{(1^k)}t^\rho, \mba^2 t^{-\rho})
\prod_{i=1}^k \left(\mbq^{2i-2} - \mba^2\right) 
\CR
&=& \mba^{-k-1} \mbt^{k} \mbq^{\frac{1}{2}k(k-1)} 
\prod_{i=1}^k \frac{\mbq^{2i-2}- \mba^2}{\mbq^i - \mbq^{-i}}
\left( \mbt^{-2} \frac{\mbq^{2k-2} - \mbq^{-2}}{\mbq - \mbq^{-1}} 
+ \frac{\mba^{2} - \mbq^{2k}}{\mbq - \mbq^{-1}} \right).
\eeqa
When $k=1$, we find
\beqa
\overline{\cal P}_{\unitbox,~\unitbox} (\mba, \mbq, \mbt) 
&=& \mbt~\frac{\mba^{-2}-1}{\mbq - \mbq^{-1}} 
\left(\mbt^{-2} \frac{1-\mbq^{-2}}{\mbq - \mbq^{-1}}
+ \frac{\mba^2 - \mbq^2}{\mbq - \mbq^{-1}} \right) \CR
&=& \frac{(-\mbt)^{-1}}{\mba^2 (1- \mbq^2)^2}
\left(\mba^4\mbt^2\mbq^2 - \mba^2(\mbt^2\mbq^4 + \mbt^2\mbq^2
-\mbq^2 +1) +\mbt^2\mbq^4 - \mbq^2 +1 \right)~. \CR
\eeqa
Up to the factor $(-\mbt)^{-1}$ or a shift of $\mbt$-grading by $-1$, 
this result agrees with \cite{GIKV},
where it was shown that for $\mba=\mbq^N$ this superpolynomial
gives rise to the Khovanov-Rozansky invariants of the Hopf link. 

It is remarkable that for any $k$ the substitution of  $\mba= \mbq^N$ reduces
the superpolynomial $\overline{\cal P}_{(1^k), \unitbox} (\mba, \mbq, \mbt)$ 
to a polynomial in $\mbq^{\pm 1}$ and $\mbt^{\pm 1}$;
\beqa
\overline{\cal P}_{(1^k),~\unitbox} (\mbq, \mbt) 
&=& \mbt^{k} \mbq^{Nk -N +\frac{1}{2}k(k-1)} 
\prod_{i=1}^k \frac{\mbq^{2i-2-2N}-1}{\mbq^i - \mbq^{-i}}
\left( \mbt^{-2} \mbq^{k-2} \frac{\mbq^{k} -\mbq^{-k}}{\mbq - \mbq^{-1}}
+ \frac{\mbq^{2N} - \mbq^{2k}}{\mbq - \mbq^{-1}}
\right) \CR
  &=& \mbq^{-2N} \mbq^{k(k-1)} (-\mbt)^{k}
\left[ N \atop k \right]_\mbq
\left( [k]_\mbq {\mathbf  q}^{N+k-2} \mbt^{-2} + [N-k]_\mbq
 {\mathbf  q}^{2N+k} \right)~, \label{yonezawa}
\eeqa
which gives homological invariants of the Hopf link with representations 
$(1^k, \unitbox)$ of ${\mathfrak sl}(N)$. 
Note that this expression vanishes for $k >N$ 
as it should be for the totally antisymmetric representation $(1^k)$ of ${\mathfrak sl}(N)$.
We have factorized the standard normalization factor $\mbq^{-2N}$ for the Hopf link.
Since both the $q$-integer and the $q$-binomial coefficient are
polynomials with positive integer coefficients, 
\eqref{yonezawa} shows that $\overline{\cal P}_{(1^k),~\unitbox} (\mbq, \mbt)$ is
a Laurent polynomial in $\mbq$ and $\mbt$ with positive integer coefficients. 
Up to an overall factor $ \mbq^{k(k-1)}(-\mbt)^{k}$, 
\eqref{yonezawa} agrees to the recent result by Yonezawa by matrix factorization \cite{Yone}.

More generally we can show that the superpolynomial 
$\overline{\cal P}_{(1^r), (1^s)} (\mba, \mbq, \mbt)$
always gives a polynomial in $\mbq^{\pm 1}, \mbt^{\pm 1}$,
when we substitute $\mba= \mbq^N$.  In fact
after the substitution we have
\beqa
\overline{\cal P}_{(1^r), (1^s)} (\mbq, \mbt) 
&=& (-1)^{r(s-1)} \mbq^{N(r-s)+s(s-1)+ \frac{1}{2}r(r-1)} \mbt^{r\cdot s}
\prod_{i=1}^r  \frac{\mbq^{2i-2-2N}-1}{\mbq^{i} - \mbq^{-i}}
e_s(q^{(1^r)}t^\rho, \mbq^{2N} t^{-\rho}) \CR
&=& (- \mbt)^{r\cdot s} \mbq^{-sN+s(s-1)+ r(r-1)}
\left[  N \atop r  \right]_\mbq
e_s(q^{(1^r)}t^\rho, \mbq^{2N} t^{-\rho})~.
\eeqa
Hence what we have to show is that $e_s(q^{(1^r)}t^\rho, \mbq^{2N} t^{-\rho})$ is a polynomial.
Since the elementary symmetric function $e_s$ is a polynomial in the power sums
$p_1, \cdots, p_s$, let us look at $p_k(q^{(1^r)}t^\rho, \mbq^{2N} t^{-\rho})$.
We find
\beqa
p_k(q^{(1^r)}t^\rho, \mbq^{2N} t^{-\rho})
&=& (\mbt\mbq)^{-2k} \frac{\mbq^{2kr} -1}{\mbq^{k} - \mbq^{-k}}
+ \frac{\mbq^{2kN}-\mbq^{2kr}}{\mbq^{k} - \mbq^{-k}} \CR
&=& \mbt^{-2k} \mbq^{k(r-2)} \frac{[kr]_\mbq}{[k]_\mbq}
+ \mbq^{k(N+r)} \frac{[k(N-r)]_\mbq}{[k]_\mbq}~,
\eeqa
which shows that they are polynomials.
Thus we see that $e_s(q^{(1^r)}t^\rho, \mbq^{2N} t^{-\rho})$ is a polynomial
in $\mbq^{\pm 1}, \mbt^{\pm 1}$.


\Section{Case of other representations}


Using the fact that the Macdonald functions and the Schur functions are related by
a base change of the space of symmetric functions, we can compute 
the homological invariants for other representations than totally anti-symmetric 
representations. The base change is rather complicated in general. But let us
illustrate the method of computation in the simplest examples. 


\subsection{$(\la,\mu)=((2),(1^s))$ case}


When $|\lambda|=2$,
the base change is
\beq
\Smat{
s_{(2)}(x) \\
s_{(1^2)}(x)
}
=
\Smat{
~1~ & \ds{\frac{t-q}{1-qt}} \\
0 &  1
}
\Smat{
P_{(2)} (x; q,t) \\
P_{(1^2)} (x; q,t) 
}.
\eeq
By transforming the Schur functions to the Macdonald functions,
we can apply a similar way of computation to the case of totally anti-symmetric representations.
We obtain
\beq
\Smat{
\frac{Z_{(2), (1^s)}(Q; q,t)}{Z_{\bullet, \bullet}(Q; q,t)}  \\
\frac{Z_{(1^2), (1^s)}(Q; q,t)}{Z_{\bullet, \bullet}(Q; q,t)}
}
=
(-1)^s t^{-\frac{1}{2}s(s-1)} 
\Smat{
~1~ & \ds{\frac{t-q}{1-qt}} \\
0 &  1
}
\Smat{
P_{(2)}(t^\rho; q,t) N_{(2),\bullet}(\tQ; q,t) 
e_s(\tQ q^{(2)} t^\rho, t^{-\rho}) \\
P_{(1^2)}(t^\rho; q,t) N_{(1^2),\bullet}(\tQ; q,t) 
e_s(\tQ q^{(1^2)} t^\rho, t^{-\rho}) 
},
\eeq
and hence
\beqa
\overline{\cal P}_{(2), (1^s)} (\mba, \mbq, \mbt) 
&=& \mba^{s+2} \mbt^{4s} \mbq^{s(s-1)}
\left(
(s_{(2)}(t^\rho) - \frac{t-q}{1-qt} s_{(1^2)}(t^\rho))N_{(2),\bullet}(\tQ; q,t) e_s(\tQ q^{(2)} t^\rho, t^{-\rho}) \right. \CR
&&~~~\left.+ \frac{t-q}{1-qt} s_{(1^2)}(t^\rho) N_{(1^2),\bullet}(\tQ; q,t) e_s(\tQ q^{(1^2)} t^\rho, t^{-\rho}) \right)~.
\eeqa
The terms proportional to $\frac{t-q}{1-qt}$ give a discrepancy to the conjecture \eqref{Takiconj}.
It holds only for totally anti-symmetric representations.
As we argued in section 3, this means that the slice invariance 
of open topological string amplitudes is broken, if the representation
attached to the topological brane is not totally anti-symmetric. 
The above formula for $s=0$ provides the following superpolynomial 
for the unknot colored by the symmetric representation;
\beq
\overline{\cal P}_{(2), \bullet} (\mba, \mbq, \mbt) 
= \mba^{-2} \mbt^{-4} \mbq^{-7} \frac{(\mba^{2} -1)}
{(\mbq - \mbq^{-1})(\mbq^2 - \mbq^{-2})}
\left(  \mbt^2 \mbq^2 (\mba^2 - \mbq^2) + \mbq^4 -1 \right)~.
\eeq
When $s=1$, we find the following superpolynomial
for the Hopf link colored by the symmetric representation and 
the fundamental representation\footnote{Examples of the superpolynomials 
provided in appendix of \cite{GIKV} are only for totally antisymmetric representations
and this is a new example of the superpolynomial for the colored Hopf link.};
\beqa
\overline{\cal P}_{(2), (1)} (\mba, \mbq, \mbt) 
&=& \mba~\mbt^{4}\frac{\mbq}{(\mbq - \mbq^{-1})(\mbq^2 - \mbq^{-2})} \CR
& &~~\left(
(\mbq^{-2} - \frac{\mbq^{-2} - \mbt^{-2}\mbq^{-2}}{1- \mbt^{-2}\mbq^{-4}})
(1- \mba^{-2})(1- \mba^{-2}\mbt^{-2}\mbq^{-2})
(\mbt^{-4}\mbq^{-3} + \frac{\mba^2 - \mbq^2}{\mbq - \mbq^{-1}}) \right. \CR
& &~~~+ \left. \frac{\mbq^{-2} - \mbt^{-2}\mbq^{-2}}{1- \mbt^{-2}\mbq^{-4}}
(1- \mba^{-2})(1-\mba^{-2}\mbq^2)(\mbt^{-2}\mbq^{-2} (\mbq + \mbq^3)
+ \frac{\mba^2 - \mbq^4}{\mbq - \mbq^{-1}})\right) \CR
&=&  \mba^{-3}\mbt^{4}\mbq^{-1}\frac{(\mba^{2} -1)}
{(\mbq - \mbq^{-1})^2(\mbq^2 - \mbq^{-2})}  \CR
& &\left((\mba^2 - \mbq^4)(\mba^2 - \mbq^2)
+\mbt^{-2}\mbq^{-2}(\mbq^2 -1)(\mbq^2 +1)^2(\mba^2 - \mbq^2) \right. \CR
& & \left.-\mbt^{-4}\mbq^{-4}(\mbq^2 -1)^2(\mbq^2 +1)(\mba^2 - \mbq^2)
+\mbt^{-6}\mbq^{-6}(\mbq^2 -1)^2(\mbq^2 +1) \right)~. \CR
\eeqa
Note that there is a cancellation of the factor $(1-qt)=(1- \mbt^{-2}\mbq^{-4})$ in the denominator. 
Finally by substituting $\mba=\mbq^N$ we can eliminate the remaining factors
in the denominator to get a polynomial in $\mbq^{\pm 1}$ and $\mbt^{\pm 1}$.
We believe this is a highly non-trivial consistency check of our formula \eqref{theorem}.


\subsection{$(\la,\mu)=((21),(1^s))$ case}


The second example is the hook representation $\lambda=(21)$ with $|\lambda|=3$.
In this case we have
\beq
\Smat{
s_{(3)}(x) \\
s_{(21)}(x) \\
s_{(1^3)}(x)
}
=
\Smat{
1 & \ds{(t-q)(1+q) \over 1-q^2t} &  \ds{(t-q)(t^2-q) \over (1-qt)(1-q^2t)} \\
~0~ &1 & \ds{\frac{(t-q)(t+1)}{1-qt^2}} \\
0 & 0 &  1
}
\Smat{
P_{(3)} (x; q,t) \\
P_{(21)} (x; q,t) \\
P_{(1^3)} (x; q,t) 
}.
\eeq
By the same manner as above we obtain the superpolynomial
\beqa
& &\overline{\cal P}_{(21), (1^s)} (\mba, \mbq, \mbt) \CR
&=& -\mba^{s+3} \mbt^{6s} \mbq^{s(s-1)}
\left(
(s_{(21)}(t^\rho) - \frac{(t-q)(t+1)}{1-qt^2} s_{(1^3)}(t^\rho))N_{(21),\bullet}(\tQ; q,t) e_s(\tQ q^{(21)} t^\rho, t^{-\rho}) \right. \CR
&&~~~\left.+ \frac{(t-q)(t+1)}{1-qt^2} s_{(1^3)}(t^\rho) N_{(1^3),\bullet}(\tQ; q,t) e_s(\tQ q^{(1^3)} t^\rho, t^{-\rho}) \right)~.
\eeqa
When $s=0$ the superpolynomial for the unknot is
\beq
\overline{\cal P}_{(21),\bullet} (\mba, \mbq, \mbt)
 = \mba^{-3} \mbt^{-4} \mbq^{-9}
 \frac{(\mba^{2} -1)(\mba^{2} - \mbq^2)}
{(\mbq - \mbq^{-1})^2(\mbq^3 - \mbq^{-3})} 
 ( \mbt^2 \mbq^2(\mba^2 - \mbq^4) + \mbq^6 - 1)~.
\eeq
When $s=1$ the superpolynomial is
\beqa
& & \overline{\cal P}_{(21),~\unitbox} (\mba, \mbq, \mbt) \CR
&=& \mba^2 \mbt^6 \frac{(1- \mba^{-2})(1-\mba^{-2} \mbq^2)}
{(\mbq - \mbq^{-1})(\mbq^2 - \mbq^{-2})(\mbq^3 - \mbq^{-3})}  \CR
& &~~~\left(
(\mbq + \mbq^{-1}- \frac{\mbq(1- \mbt^{-2})(1+\mbq^{-2})}{1- \mbt^{-2}\mbq^{-6}})
(1- \mba^{-2} \mbt^{-2} \mbq^{-2})( \mbt^{-4} \mbq^{-3} 
+ \mbt^{-2} \mbq + \frac{\mba^2 - \mbq^4}{\mbq - \mbq^{-1}} ) \right. \CR
& &~~~\left.+\frac{\mbq(1- \mbt^{-2})(1+\mbq^{-2})}{1- \mbt^{-2}\mbq^{-6}})
(1- \mba^{-2} \mbq^4)(\mbt^{-2}(\mbq^{-1} + \mbq + \mbq^3) + \frac{\mba^2 - \mbq^6}{\mbq - \mbq^{-1}} \right) \CR
&=&  \mba^{-4} \frac{(\mba^{2} -1 )(\mba^{2} -\mbq^2)}
{(\mbq - \mbq^{-1})^3(\mbq^3 - \mbq^{-3})} 
(\mbt^6P_6(\mba, \mbq) +  \mbt^4\mbq^{-2} P_4(\mba, \mbq) 
+ \mbt^2\mbq^{-4} P_2(\mba, \mbq)  + \mbq^{-6} P_0(\mba, \mbq))~, \CR
\eeqa
where
\beqa
P_6(\mba, \mbq) &=& (\mba^2 - \mbq^4)(\mba^2 - \mbq^6)~, \CR
P_4(\mba, \mbq) &=& (\mba^2 - \mbq^4)(\mbq^2 -1)(\mbq^2 +1)(\mbq^2 + \mbq +1)(\mbq^2 - \mbq +1)~, \\
P_2(\mba, \mbq) &=& - (\mba^2 - \mbq^2 -\mbq^4) (\mbq^2 -1)^2 (\mbq^2 + \mbq +1)(\mbq^2 - \mbq +1)~, \CR
P_0(\mba, \mbq) &=&   (\mbq^2 -1)^2 (\mbq^2 + \mbq +1)(\mbq^2 - \mbq +1)~. \nonumber
\eeqa
We can again confirm a cancellation of the factor $(1-qt^2)=(1- \mbt^{-2}\mbq^{-6})$ in the denominator
and the fact that the superpolynomial reduces to a polynomial in 
$\mbq^{\pm 1}$ and $\mbt^{\pm 1}$ when $\mba = \mbq^N$.


\subsection{$(\la,\mu)=(\la,1^s)$ case}


Generalizing the above computations,
we can show the \lq\lq finiteness\rq\rq\ of the superpolynomial
for $(\la,\mu)=(\la,1^s)$. As a Laurent polynomial in $\mba$ 
the degree of $\overline{\cal P}_{\la, (1^s)} (\mba, \mbq, \mbt)$ ranges
from $-|\la|-s$ to $\la+s$. When we substitute $\mba=\mbq^N$ with
a fixed $N$, $\overline{\cal P}_{\la, (1^s)} (\mbq^N, \mbq, \mbt)$ vanishes
if $N < \max(|\la|,s)$. By the relation \eqref{superpol} 
these properties follow from the propositions below. 
Since the set of the Macdonald functions $\{P_\la (x;q,t)\}_{|\la |=d}$
is a basis of the symmetric functions of homogeneous degree $d$, 
we can write the Schur function $s_{\la}(x)$
by the Macdonald functions as
\be
s_{\la}(x) = 
\sum_{\mu \atop |\mu|=|\la |}
U_{\la,\mu}(q,t) P_\mu (x;q,t), 
\label{basechange}
\ee
where $U_{\la ,\mu}(q,t)$ is a rational function. 
Then we have
\\
{\bf Proposition.}~{\it  
For $|t|<1$,
\ba
{
Z_{\la ,1^s}(Q;q,t) 
\over 
Z_{\bullet,\bullet}(Q;q,t) 
}
&=&
g_{1^s}
\sum_{\mu \atop |\mu|=|\la |}
U_{\la,\mu}(q,t) 
g_\mu 
P_\mu (vQt^{\rho},t^{-\rho};q,t)
e_s(vQq^\mu  t^\rho, t^{-\rho}),
\cr
g_\la
&:=&
\prod_{s\in\la }(-1)q^{a(s)}t^{-\ell(s)},
\ea
which is a polynomial of degree $|\la | + s$ in $Q$. 
}


\proof
From \eqref{basechange}
\be
Z_{\la ,1^s}(Q;q,t) 
=
\sum_\eta (-Q)^{|\eta|}
P_{\eta^\vee}(q^{\rho};t,q)
P_{\eta}(t^{\rho};q,t)
s_{\la}(q^{\eta}t^{\rho})
s_{1^s}(q^{\eta}t^{\rho})
=
\sum_{\mu \atop |\mu|=|\la |}
U_{\la,\mu}
\widetilde Z_{\mu ,1^s}(Q;q,t), 
\ee
where
\be
\widetilde Z_{\mu ,1^s}(Q;q,t) 
:=
\sum_\eta (-Q)^{|\eta|}
P_{\eta^\vee}(q^{\rho};t,q)
P_{\eta}(t^{\rho};q,t)
P_{\mu}(q^{\eta}t^{\rho};q,t)
s_{1^s}(q^{\eta}t^{\rho}).
\ee
Then by 
(\ref{eq:PPsymm}) and 
(\ref{eq:Pdual}),
\be
\widetilde Z_{\mu ,1^s}(Q;q,t) 
=
\sum_\eta {(vQ)^{|\eta|}\over \langle P_\eta \vert P_\eta \rangle_{q,t} }
P_{\eta}(t^{-\rho};q,t)
P_{\mu}(t^{\rho};q,t)
P_{\eta}(q^\mu t^{\rho};q,t)
s_{1^s}(q^{\eta}t^{\rho}).
\ee
Since
$s_{1^s}(q^{\eta}t^{\rho}) = e_s(q^{\eta}t^{\rho})$,
by
(\ref{eq:EigenHN}) and 
(\ref{eq:HNPiPi}),
\ba
\widetilde Z_{\mu ,1^s}(Q;q,t) 
&=&
P_{\mu}(t^{\rho};q,t)
\sum_\eta {(vQ)^{|\eta|}\over \langle P_\eta \vert P_\eta \rangle_{q,t} }
P_{\eta}(q^\mu t^{\rho};q,t)
H^s P_{\eta}(x;q,t)\vert_{x=t^{-\rho}}.
\cr
&=&
P_{\mu}(t^{\rho};q,t)
H^s
\Pi(x,vQq^{\mu }t^{\rho};q,t)
\vert_{x=t^{-\rho}}
\cr
&=&
P_{\mu}(t^{\rho};q,t)
(-1)^s t^{-{s(s-1)\over 2}}
e_s(vQq^\mu t^\rho,t^{-\rho})
\Pi(t^{-\rho},vQq^{\mu }t^{\rho};q,t).
\ea
But by \eqref{eq:NekIIIp}
and ((5.20) in \cite{AK2})
\be
P_\mu (t^{\rho};q,t)
N_{\mu , \bullet}(vQ;q,t)
=
P_\mu (vQt^{\rho},t^{-\rho};q,t)
v^{-|\mu |} f_\mu (q,t),
\ee
with the framing factor $f_\mu (q,t)$ defined by \eqref{eq:framing},
we have
\be
{
\widetilde Z_{\mu ,1^s}(Q;q,t) 
\over 
Z_{\bullet,\bullet}(Q;q,t) 
}
=
(-1)^s t^{-{s(s-1)\over 2}}
v^{-|\mu |} 
P_\mu (vQt^{\rho},t^{-\rho};q,t)
e_s(vQq^\mu  t^\rho, t^{-\rho})
f_\mu (q,t).
\ee
\qed


Note that, for $N\in\bZ$ and $N\geq\ell(\la )$,
\be
p_n(q^\lambda t^\rho, t^{-N-\rho} ) 
=
\sum_{i=1}^{\ell(\la)} (q^{n\lambda_i}-1)t^{n(\ha-i)}
+ {1 - t^{-Nn} \over t^{n\over 2} - t^{-{n\over 2}} }
=
\sum_{i=1}^N q^{n\lambda_i}t^{n(\ha-i)}.
\ee
Hence,
$P_\mu (q^\la t^\rho, t^{-N-\rho} )$
is the Macdonald polynomial in $N$ variables
$\{ q^{\la_i} t^{\ha-i} \}_{1\leq i \leq N}$
and vanishes for $\ell(\la )\leq N <\ell(\mu)$.
Therefore when $vQ=t^N$, we have 
\\
{\bf Proposition.}~{\it
If $N\in\bZ_{\geq 0}$ and $|t|<1$,
\ba
{
Z_{\lambda ,1^s}(v^{-1}t^N;q,t) 
\over 
Z_{\bullet,\bullet}(v^{-1}t^N;q,t) 
}
&=&
g_{1^s}
\sum_{\mu \atop |\mu|=|\la |}
U_{\la,\mu} g_\mu
P_\mu (t^{N+\rho},t^{-\rho};q,t)
P_{1^s}(q^\mu  t^{N+\rho}, t^{-\rho};q,t)
\cr
&=&
g_{1^s}
P_{1^s}(t^{N+\rho},t^{-\rho};q,t)
\sum_{\mu \atop |\mu|=|\la |}
U_{\la,\mu} g_\mu
P_\mu(q^{(1^s)} t^{N+\rho}, t^{-\rho};q,t),
\label{eq:LaAntiQtN}%
\ea
which vanishes for $0\leq N < \max (|\lambda|,s)$. 
}


\proof
If $0\leq N< |\mu|$, $P_\mu (t^{\rho},t^{-N-\rho};q,t)=0$.
If $|\mu|\leq N< s$, $P_{1^s}(q^\mu  t^\rho, t^{-N-\rho};q,t)=0$.
Similarly, if $0\leq N< s$, $P_{1^s}(t^{\rho},t^{-N-\rho};q,t)=0$.
If $s\leq N< |\mu|$, $P_\mu(q^{(1^s)} t^\rho, t^{-N-\rho};q,t)=0$.
Hence the middle and the right hand side of (\ref{eq:LaAntiQtN})
vanish for $0\leq N < \max (|\mu|,s)$.

On the other hand, if $N \geq \max (|\mu|,s)$,
by the finite $N$ version of (\ref{eq:PPsymm}),
\be
P_{1^s} (t^{\rho},t^{-N-\rho};q,t) 
P_\mu(q^{1^s } t^\rho, t^{-N-\rho};q,t)
=
P_\mu (t^{\rho},t^{-N-\rho};q,t) 
P_{1^s} (q^\mu  t^\rho, t^{-N-\rho};q,t),
\ee
we have the proposition.
\qed

The condition $|t|<1$ in the last two propositions can be eliminated \cite{AK4}.


Finally we should make a remark on the fact that
the transition function $U_{\la ,\mu}(q,t)$ in \eqref{basechange}
is a rational function in $q$ and $t$. 
It is not obvious that the formulae in the above propositions 
are in fact polynomials in $q$ and $t$. 
However, we have checked the following conjecture up to $d=7$
by direct calculation;
\\
{\bf Conjecture.}~{\it
For $|\lambda|=d$,
$
\sum_{\mu, |\mu|=d} 
U_{\la ,\mu}
g_\mu
(U^{-1})_{\mu, \nu}
$
is a polynomial 
of degree $d(d-1)/2$ in $q$ and 
of degree $d(d-1)/2$ in $t^{-1}$.
}

Under the above conjecture,
if $N\in\bZ_{\geq 0}$,
\ba
{
Z_{\la ,1^s}(v^{-1}t^N;q,t) 
\over
Z_{\bullet,\bullet}(v^{-1}t^N;q,t) 
}
&=&
g_{1^s}
e_s(t^{N+\rho},t^{-\rho})
\sum_{\mu,\nu\atop |\mu|=|\nu|=|\la |} 
U_{\la ,\mu}
g_\mu
(U^{-1})_{\mu, \nu}
s_\nu(q^{(1^s)} t^{N+\rho}, t^{-\rho}),
\ea
is a polynomial in $q$ and $t$.


\section*{Acknowledgments}


We would like to thank M.~Taki and Y.~Yonezawa for discussions. 
We also thank T.~Eguchi and H.~Nakajima for helpful comments and
correspondence. 
This work is partially supported by the Grant-in-Aid for Nagoya
University Global COE Program, "Quest for Fundamental Principles in the
Universe: from Particles to the Solar System and the Cosmos", from the Ministry
of Education, Culture, Sports, Science and Technology of Japan.
The present work is also supported in part by Daiko Foundation. 
The work of H.K.  is supported in part by Grant-in-Aid for Scientific Research
[\#19654007] from the Japan Ministry of Education, Culture, Sports, Science and Technology

\Section*{Appendix: Five-dimensional $U(1)^N$ theory}
\renewcommand{\theequation}{A.\arabic{equation}}\setcounter{equation}{0}
\renewcommand{\thesubsection}{A.\arabic{subsection}}\setcounter{subsection}{0}


Here we generalize the argument in section 2
to the five-dimensional $U(1)^N$ theory.
We compare two partition functions 
associated with the diagrams in \fgref{fg:UN}.
Since the difference is just the choice of the preferred directions,
we expect that they coincide.

\FigUN


Let
\ba
Z_L
&:=&  
\sum_{ \{\Ya\la \a \} }
\prod_{\a =1}^N
\Ciio{\bullet}{\Ya\la {2\a -2}}{\Ya\la {2\a -1}}{q}{t}
\Cooi{\bullet}{\Ya\la {2\a   }}{\Ya\la {2\a -1}}{q}{t}
\prod_{\a =1}^{2N} 
Q_{\a }^{|\Ya\la \a  |}
\cr
&=& 
\sum_{ \{\Ya\la \a \} }
\prod_{\a =1}^N
P_{\Ya \la {2\a -2}}(t^{\rho} ;q,t) 
P_{\Ya \la {2\a -1}}(q^{\Ya\la {2\a -2}} t^{\rho} ;q,t) 
\cr
&&\hskip30pt
\times
P_{\Yav\la {2\a   }}(-q^\rho ;t,q) 
P_{\Yav\la {2\a -1}}(-t^{\Yav\la {2\a }} q^{\rho} ;t,q) 
\prod_{\a =1}^{2N} 
Q_{\a }^{|\Ya\la \a  |},
\ea
with 
$\Ya\la 0 = \Ya\la {2N}$ and $\Ya\sigma  0 = \Ya\sigma  {2N}$.
The summations over the partitions with odd suffices
${\Ya\la 1}$, ${\Ya\la 3}$, ${\Ya\la 5},\cdots$, 
are performed by the Cauchy formula
\eqref{eq:skewCauchy} for $\mu=\nu=\bullet$.
From the specialization formula:
\be
P_\lambda(t^\rho;q,t) P_{\lambda^\vee} (-q^\rho;t,q)
=
{ v^{-|\lambda|}\over N_{\lambda \lambda}(1;q,t)},
\label{eq:specialPP}
\ee
with \eqref{eq:NekIIp}
we have
\be
Z_L
=
\sum_{ \{\Ya\la {2\a }\} }
\prod_{\a =1}^N
{\(v^{-1}Q_{2\a }\)^{|\Ya\la {2\a }|}
\over 
N_{{\Ya\la {2\a }}{\Ya\la {2\a }} }(1;q,t)}
\Pi_0
(q^{\Ya\la {2\a -2}} t^{\rho} , -Q_{2\a -1}t^{\Yav\la {2\a }} q^{\rho} ).
\ee


If we separate out the perturbative part
$
Z_L^{\rm pert} 
:=
Z_L(Q_{2\a }=0)
=
\prod_{\a =1}^N
\Pi_0
(t^{\rho} , -Q_{2\a -1}q^{\rho} )
$,
then from
\eqref{eq:NekIIp},
$
Z_L^{\rm inst}
:=
{Z_L / Z_L^{\rm pert} }
$
is
\be
Z_L^{\rm inst}
=
\sum_{ \{\Ya\la {2\a }\} }
\prod_{\a =1}^N
{\tQ_{2\a }^{|\Ya\la {2\a }|}\over 
N_{{\Ya\la {2\a }}{\Ya\la {2\a }} }(1;q,t)}
N_{{\Ya\la {2\a -2}}{\Ya\la {2\a }} }(\tQ_{2\a -1};q,t).
\ee


On the other hand, let
\ba
Z_R
&:=&  
\sum_{ \{\Ya\la \a \} }
\prod_{\a =1}^N
\Coii{\Ya\la {2\a -1}}{\bullet}{\Ya\la {2\a -2}}{q}{t}
\Cioo{\Ya\la {2\a -1}}{\bullet}{\Ya\la {2\a   }}{q}{t}
\prod_{\a =1}^{2N} 
Q_{\a }^{|\Ya\la \a  |}
\cr
&=& 
\sum_{ \{\Ya\la \a \} }
\prod_{\a =1}^N
\sum_{\Ya\sigma  {2\a -1}}
\iota P_{{\Yav\la {2\a -1}}/{\Yav\sigma  {2\a -1}}}
(-q^{\rho} ;t,q) 
P_{\Ya\la {2\a -2}/\Ya\sigma  {2\a -1}} 
(t^\rho ;q,t) 
\cr
&&\times
\sum_{\Ya\sigma  {2\a }}
P_{{\Yav\la {2\a }}/{\Yav\sigma  {2\a }}}
(-q^{\rho} ;t,q) 
\iota P_{\Ya\la {2\a -1}/\Ya\sigma  {2\a }} 
(t^\rho ;q,t) 
\prod_{\a =1}^N
v^{|\Ya\sigma  {2\a -1}|-|\Ya\sigma  {2\a }|}
\prod_{\a =1}^{2N} 
Q_{\a }^{|\Ya\la \a  |},
\ea
with 
$\Ya\la 0 = \Ya\la {2N}$ and $\Ya\sigma  0 = \Ya\sigma  {2N}$.
%
%
The following trace formula is useful for calculating $Z_R$.
%
\\
{\bf Proposition.}\cite{AK2}~{\it 
For $N\in\bN$, 
let $x^\a =x^{\a +2N}$'s  be sets of variables,
$\Ya\lambda \a $'s be the Young diagrams,
$\Ya\lambda 0 = \Ya\lambda{2N}$,
$c_{\a ,\a +1}=c_{\a +2N,\a +1+2N}\in\bC$, 
$c_{\a ,\b }:=\prod_{n=\a }^{\b -1} c_{\a ,\a +1}$ and
$c:=c_{1,1+2N}=\prod_{\a =1}^{2N}c_{\a ,\a +1}$.
If $|c|<1$, then
\ba
&&\hskip -30pt
\sum_{\{\Ya\lambda 1,\Ya\lambda 2,\cdots,\Ya\lambda{2N}\}}
\prod_{\a =1}^N
P_{{\Ya \lambda{2\a -2}}/{\Ya \lambda{2\a -1}} } (x^{2\a -1};q,t) 
P_{{\Yav\lambda{2\a   }}/{\Yav\lambda{2\a -1}} } (x^{2\a   };t,q) 
\cdot
\prod_{\a =1}^{2N}
c_{\a ,\a +1}^{|\Ya\lambda \a |}
\cr
&=&
\prod_{n\geq 0} {1\over 1-c^{n+1}}
\prod_{\a =1}^N \prod_{\b =0}^{N-1}
\Pi_0( x^{2\a }, \ c_{2\a ,2\a +2\b +1} c^n  x^{2\a +2\b +1})
\cr
&=&
\exp\left\{-\sum_{n>0}{1\over n}
{1\over 1-c^n} 
\left\{
\sum_{\a =1}^N \sum_{\b =0}^{N-1}
c_{2\a ,2\a +2\b +1}^n p_n(x^{2\a }) p_n(-x^{2\a +2\b +1})
-c^n
\right\}\right\}.
\label{eq:TraceFormula}
\ea
}

Let $(c_{4\a -3,4\a -2},c_{4\a -2,4\a -1}c_{4\a -1,4\a }c_{4\a ,4\a +1})
:=(v, Q_{2\a -1},v^{-1},Q_{2\a })$ 
with $v :=\left({q/t}\right)^{\ha}$
and
$(x^{4\a -3},x^{4\a -2},x^{4\a -1},x^{4\a })
:=(t^\rho, -\iota q^\rho,\iota t^\rho,-q^\rho)$,
then we have  
$c=\prod_{\a =1}^{2N} Q_\a $,
\be
c_{2\a ,2\a +2\b +1} = 
v^{
(-1)^\a {1-(-1)^\b  \over 2}
}
\prod_{\g = 0 }^\b Q_{\a +\g },
\ee
with
$Q_{\a +2N} := Q_\a $ 
and 
\be
p_n(x^{2\a }) p_n(-x^{2\a +2\b +1}) 
=
{(-1)^\b 
\over 
\(q^{{n\over 2}}-q^{-{n\over 2}}\)
\(t^{{n\over 2}}-t^{-{n\over 2}}\)
}.
\ee
Therefore, from \eqref{eq:TraceFormula},
it follows that
\be
Z_R
=
\exp\left\{
-\sum_{n>0} {1\over n} {1\over 1-c^n}
\left\{{
g(\{Q_\a ^n\};q^n,t^n)
\over 
\(t^{{n\over 2}}-t^{-{n\over 2}}\)
\(q^{{n\over 2}}-q^{-{n\over 2}}\)
}-c^n
\right\}
\right\},
\ee
with
\ba
g(\{Q_\a \};q,t)
&:=&
\sum_{\a =1}^{2N}
\sum_{\b =0}^{2N-1}
(-1)^\b  
v^{(-1)^\a  { 1-(-1)^\b \over 2} }
\prod_{\g =0}^{\b }
Q_{\a +\g }
\cr
&=&
\sum_{\a =1}^{2N}
v^{(-1)^\a }
\sum_{\b =0}^{2N-1}
(-1)^\b  
\prod_{\g =0}^{\b }
\tQ_{\a +\g }
\cr
&=&
\sum_{\a =1}^{2N}
v^{(-1)^\a }
\tQ_{\a }
(1-\tQ_{\a +1}
(1-\tQ_{\a +2}
(
\cdots
(1-\tQ_{\a +2N-1})
\cdots 
))).
\ea
Here
$
\tQ_\a := v^{(-1)^{\a +1}}Q_\a $.


If we separate out the perturbative part
\be
Z_R^{\rm pert} 
:=
Z_R(Q_{2\a }=0)
=
\exp\left\{
-\sum_{n>0} {1\over n}
{
\sum_{\a =1}^N Q_{2\a -1}^n
\over 
\(t^{{n\over 2}}-t^{-{n\over 2}}\)
\(q^{{n\over 2}}-q^{-{n\over 2}}\)
}
\right\},
\ee
then
$
Z_R^{\rm inst}
:=
{Z_R / Z_R^{\rm pert} }
$
is
\be
Z_R^{\rm inst}
=
\exp\left\{
-\sum_{n>0} {1\over n} {1\over 1-c^n}
\left\{{
g^{\rm inst}(\{Q_\a ^n\};q^n,t^n)
\over 
\(t^{{n\over 2}}-t^{-{n\over 2}}\)
\(q^{{n\over 2}}-q^{-{n\over 2}}\)
}-c^n
\right\}
\right\},
\ee
with
\ba
&&\hskip-10pt
g^{\rm inst}(\{Q_\a \};q,t)
:=
\sum_{\a =1}^{N}
\(
v
\sum_{\b =0}^{2N-1}
(-1)^\b  
\prod_{\g =0}^{\b }
\tQ_{2\a +\g }
+
v^{-1}
\sum_{\b =1}^{2N}
(-1)^\b  
\prod_{\g =0}^{\b }
\tQ_{2\a -1+\g }
\)
\cr
&=&
\sum_{\a =1}^{N}
(v-v^{-1}\tQ_{2\a -1})
\tQ_{2\a }
(1-\tQ_{2\a +1}
(1-\tQ_{2\a +2}
(
\cdots
(1-\tQ_{2\a +2N-1})
\cdots 
))).
\ea


Since $Z_L^{\rm pert}= Z_R^{\rm pert}$, 
the slice invariance 
$Z_L^{\rm inst} = Z_R^{\rm inst}$
is equivalent to the the following conjecture:
%
\\
{\bf Conjecture.}~{\it
\be
\sum_{ \{\Ya\la {2\a }\} } 
\prod_{\a =1}^N Q_{2\a }^{|\Ya\la {2\a }|}
{
N_{\Ya\la {2\a }\Ya\la {2\a +2}}(Q_{2\a +1};q,t)
\over
N_{\Ya\la {2\a }\Ya\la {2\a }}(1;q,t)
}
=
\exp\left\{
\sum_{n>0} {1\over n} {1\over 1-c^n}
\left\{{
f(\{Q_\a ^n\};q^n,t^n)
\over 
(1-t^n)(1-q^{-n})}+c^n
\right\}
\right\},
\ee
with
$c:=\prod_{\a =1}^{2N} Q_\a $, 
$Q_{\a +2N} := Q_\a $ 
and 
\ba
&&\hskip-10pt
f(\{Q_\a \};q,t)
:=
\sum_{\a =1}^{N}
\(
\sum_{\b =0}^{2N-1}
(-1)^\b  
\prod_{\g =0}^{\b }
Q_{2\a +\g }
+
{t\over q }
\sum_{\b =1}^{2N}
(-1)^\b  
\prod_{\g =0}^{\b }
Q_{2\a -1+\g }
\)
\cr
&=&
\sum_{\a =1}^N
\(1-{t\over q}Q_{2\a -1} \)
Q_{2\a }
(1-Q_{2\a +1}
(1-Q_{2\a +2}
(1-Q_{2\a +3}\cdots
(1-Q_{2\a +2N-1}
)
\cdots )
)). \cr
&&
\ea
}
Computer calculations support this conjecture.
We have checked this by Maple for $N\leq 3$ and $|\Ya\la {2\a }|\leq 5$.


For $N=1$, $Q_1=Q$ and $Q_2=\Lambda$, 
then $Z_L$ and $Z_R$ 
are equal to those in section 2
and the above conjecture reduces to
\eqref{eq:conjecture}.


When $N=2$ and $Q_3=Q_4=0$, the conjecture reduces to
\ba
& &
\exp \left\{
\sum_{n>0} \frac{1}{n}\frac{ t^n Q_1^n Q_2^n - q^n Q_2^n}{(1-t^n)(1-q^n)}
\right\}
=
\sum_\la Q_2^{|\la|} \frac{N_{\bullet \la}(Q_1; q,t)}{N_{\la, \la}(1;q,t)} 
\cr
& &~~~=
\sum_\la Q_2^{|\la|} \frac{\prod_{(i,j) \in \la} (1- Q_1 q^{-\la_i +j-1} t^{i})}
{\prod_{s \in \la}(1-q^{a(s)} t^{\ell(s)+1})(1- q^{-a(s)-1} t^{-\ell(s)})}~.
\ea
From the specialization formula \eqref{eq:LargeNPrincipalSpecialization}
or \eqref{eq:specialPP},
we find that this coincides with the identity of the slice invariance
discussed in subsection 5.9 of \cite{IKV}.
It is interesting that the slice independence implies the same
condition in spite of the difference of our refinement of the topological
vertex and the refined topological vertex proposed in \cite{IKV}. 
Thus the above conjecture generalizes that in \cite{IKV}.





\begin{thebibliography}{99}


\bibitem{AMV} M.~Aganagic,  M.~Mari\~no  and C.~Vafa, 
``All Loop Topological String Amplitudes From Chern-Simons Theory,''
{\it Commun. Math. Phys.} {\bf 247} (2004) 467, {\tt arXiv:hep-th/0206164}.

\bibitem{AKMV} M.~Aganagic,  A.~Klemm, M.~Mari\~no and C.~Vafa, ``The Topological Vertex,''
{\it Commun. Math. Phys. } {\bf 254} (2005) 425, {\tt arXiv:hep-th/0305132}.

\bibitem{ORV} A.~Okounkov, N.~Reshetikhin and C.~Vafa,
``Quantum Calabi-Yau and Classical Crystals,'' 
in: {\it The unity of mathematics,} Progr. Math. {\bf 244}, Birkh\"auser,  (2006),  597,
{\tt arXiv:hep-th/0309208}.

\bibitem{Nek} N.~Nekrasov, ``Seiberg-Witten Prepotential from Instanton Counting,'' 
{\it Adv. Theor. Math. Phys. } {\bf 7} (2003) 831, {\tt arXiv:hep-th/0206161}.

\bibitem{NO}
N.~Nekrasov and A.~Okounkov,
``Seiberg-Witten Theory and Random Partitions,'' 
in: {\it The unity of mathematics,} Progr. Math. {\bf 244}, Birkh\"auser, (2006), 525,
{\tt arXiv:hep-th/0306238}.

\bibitem{NY}
H.~Nakajima and K.~Yoshioka,
``Instanton Counting on Blowup I,''  {\it Invent. Math.} {\bf 162} (2005) 313,
{\tt arXiv:math.AG/0306198}.

\bibitem{HIV} T.~Hollowood, A.~Iqbal and C.~Vafa,
``Matrix Models, Geometric Engineering and Elliptic Genera,"
{\it JHEP} {\bf 0803}  (2008) 069, {\tt arXiv:hep-th/0310272}.

\bibitem{AK1}
H.~Awata and H.~Kanno, 
``Instanton counting, Macdonald function and the moduli space of $D$-branes,''
{\it JHEP}  {\bf 0505} (2005) 039, {\tt arXiv:hep-th/0502061}. 


\bibitem{Taki} M.~Taki,
``Refined Topological Vertex and Instanton Counting,''
{\it JHEP}  {\bf 0803} (2008) 048, 
{\tt arXiv:0710.1776[hep-th]}.

\bibitem{AK2}
H.~Awata and H.~Kanno, 
``Refined BPS state counting from Nekrasov's formula and
Macdonald functions,'' 
{\it Int. J. Mod. Phys.} {\bf A24} (2009) 2253, 
{\tt arXiv:0805.0191 [hep-th] }. 


\bibitem{IKV} A.~Iqbal, C.~Koz\c{c}az and C. Vafa,
``The Refined Topological Vertex,'' 
{\it JHEP}  {\bf 0910} (2009) 069,
{\tt arXiv:hep-th/0701156}.

\bibitem{GIKV} S.~Gukov, A.~Iqbal, C.~Koz\c{c}az and C. Vafa,
``Link Homologies and the Refined Topological Veretx,''
{\it Commun. Math. Phys. } {\bf 298} (2010) 757,
{\tt arXiv:0705.1368[hep-th]}.

\bibitem{IK-P1} A.~Iqbal and A.-K. Kashani-Poor,
``Instanton Counting and Chern-Simons Theory,''
{\it Adv. Theor. Math. Phys.} {\bf 7} (2003) 457, {\tt arXiv:hep-th/0212279}.

\bibitem{IK-P2} A.~Iqbal and A.-K. Kashani-Poor,
``$SU(N)$ Geometries and Topological String Amplitudes,''
{\it Adv. Theor. Math. Phys.} {\bf 10} (2006) 1, {\tt arXiv:hep-th/0306032}.

\bibitem{EK1} T.~Eguchi and H.~Kanno,
``Topological Strings and Nekrasov's Formulas,"
JHEP {\bf 0312}  (2003) 006, {\tt arXiv:hep-th/0310235}.

\bibitem{EK2} T.~Eguchi and H.~Kanno,
``Geometric transitions, Chern-Simons theory and Veneziano type amplitudes,''
{\it Phys. Lett.} {\bf B585}  (2004) 163, {\tt arXiv:hep-th/0312234}.


\bibitem{IKS}
A.~Iqbal, C.~Koz\c{c}az and K. Shabbir,
``Refined Topological Vertex, Cylindric Partitions and $U(1)$ Adjoint Theory,''
{\it Nucl. Phys.} {\bf B838}  (2010) 422,
{\tt arXiv:0803.2260[hep-th]}.


\bibitem{PS} R.~Poghossian and M.~Samsonyan,
``Instantons and the 5D U(1) gauge theory with extra adjoint,"
{\it J. Phys.} {\bf A42}  (2009) 304024,
{\tt arXiv:0804.3564 [hep-th]}.

\bibitem{EGL}
G.~Ellingsrud, L.~G\"ottsche and M.~Lehn, 
``On the Cobordism Class of the Hilbert Scheme of a Surface,"
{\tt arXiv:math.AG/9904095}.


\bibitem{DMVV}
R.~Dijkgraaf, G.~Moore, E.~Verlinde and H.~Verlinde,
``Elliptic genera of symmetric products and second quantized strings,"
{\it Commun. Math. Phys.} {\bf 185} (1997) 197,
{\tt arXiv:hep-th/9608096}.

\bibitem{Mac} I.G.~Macdonald, 
{\it Symmetric functions and Hall polynomials}, Second Edition, 
Oxford University Press, 1995. 


\bibitem{Taki2} M.~Taki,
``Flop Invariance of Refined Topological Vertex and Link Homologies,''
{\tt arXiv:0805.0336[hep-th]}.

\bibitem{Yone} Y.~Yonezawa,
``Quantum ($\mathfrak{sl}_n$,~$\land^mV_n$) link invariant and matrix factorizations,''
{\it Nagoya. J. Math.} {\bf 204} (2011) 69.


\bibitem{LLZ}
J.~Li, K.~Liu and J.~Zhou, 
``Topological String Partition Functions as Equivariant Indices,''
{\tt arXiv:math.AG/0412089}.


\bibitem{BL} L.~Borisov and A.~Libgober, 
``Elliptic genera of singular varieties,"
{\it Duke Math. J.} {\bf 116} (2003) 319, 
{\tt arXiv:math.AG/0007108}.


\bibitem{Wae} R.~Waelder,
``Equivariant Elliptic Genera and Local
MaKay Correspondence,"
{\it Asian J. Math.} {\bf 12} (2008) 251, 
{\tt arXiv:math.AG/0701336}.



\bibitem{GSV}
S.~Gukov, A.~Schwarz and C.~Vafa,
``Khovanov-Rozansky Homology and Topological Strings,''
{\it Lett. Math. Phys.} {\bf 74} (2005) 53, 
{\tt arXiv:hep-th/0412243}.





\bibitem{DGR}
N.M.~Dunfield, S.~Gukov and J.~Rasmussen, 
``The Superpolynomial for Knot Homologies,''
{\it Experiment. Math.} {\bf 15} (2006) 129, 
{\tt arXiv:math.GT/0505662}.

\bibitem{GW}
S.~Gukov and J.~Walcher,
``Matrix Factorizations and Kauffman Homology,''
{\tt arXiv:hep-th/0512298}.


\bibitem{Wit} E.~Witten, ``Chern-Simons gauge theory as a string theory,"
{\it Prog. Math.} {\bf 133} (1995) 637,  {\tt arXiv:hep-th/9207094}.


\bibitem{GV} R.~Gopakumar and C.~Vafa, 
``On the Gauge Theory/Geometry Correspondence,''
{\it Adv. Theor. Math. Phys.} {\bf 3} (1999) 1415,
{\tt arXiv:hep-th/9811131}.


\bibitem{OV} H.~Ooguri and C.~Vafa,
``Knot Invariants and Topological Strings,"
{\it Nucl. Phys.} {\bf B577} (2000) 419, {\tt arXiv:hep-th/9912123}.


\bibitem{LMV} J.M.F.~Labastida, M.~Mari\~no  and C.~Vafa,
``Knots, links and branes at large $N$," 
{\it JHEP} {\bf 0011}  (2000) 007, {\tt arXiv:hep-th/0010102}.

\bibitem{MV} M.~Mari\~no  and C.~Vafa, 
``Framed Knots at large $N$," {\tt arXiv:hep-th/0108064}.



\bibitem{AKOS}
H. Awata, H. Kubo, S. Odake and J. Shiraishi,
``Quantum $W_N$ Algebras and Macdonald Polynomials,"
 {\it Comm. Math. Phys.} {\bf 179}  (1996) 401,
{\tt arXiv:q-alg/9508011}.


\bibitem{AOS}
  H.~Awata, S.~Odake and J.~Shiraishi,
  ``Integral Representations of the Macdonald Symmetric Polynomials,"
   {\it Comm. Math. Phys.} {\bf 179} (1996) 647,
  {\tt arXiv:q-alg/9506006}.


\bibitem{AK4}
H.~Awata and H.~Kanno, 
``Macdonald Operators and Homological Invariants of the 
Colored Hopf Link,"
{\it J. Phys.} {\bf A44}  (2011) 375201,
{\tt arXiv:0910.0083 [math.QA]}.





\end{thebibliography}
\end{document}